\pdfoutput=1
\documentclass[aip,jmp,reprint,superscriptaddress]{revtex4-1}
\usepackage{blindtext}
\usepackage{amsmath,amssymb,xspace,graphicx}

\usepackage[version=3]{mhchem} % Formula subscripts using \ce{}

\begin{document}

%\title[\texttt{} tannin-controlled micelles and fibrils]{Tannin-controlled micelles and fibrils of $\kappa$-casein}
\title{Tannin-controlled micelles and fibrils of $\kappa$-casein}

\author{Wei Ma}
\affiliation{PASTEUR, D\'epartement de Chimie, \'Ecole Normale Sup\'erieure, PSL University, Sorbonne Université, CNRS,
24, rue Lhomond, 
F-75005 Paris, France}
\affiliation{State Key Laboratory for Mechanical Behavior of Materials, Xi’an Jiaotong University, Xi’an 710049, China}
\author{Christophe Tribet}
\affiliation{PASTEUR, D\'epartement de Chimie, \'Ecole Normale Sup\'erieure, PSL University, Sorbonne Université, CNRS,
24, rue Lhomond, 
F-75005 Paris, France}
\author{Sylvain Guyot}
\affiliation{INRA UR1268 BIA – Polyphenols, Reactivity, Processes, F-35653 Le Rheu, France}
\author{Dra\v{z}en Zanchi} 
\affiliation{PASTEUR, D\'epartement de Chimie, \'Ecole Normale Sup\'erieure, PSL University, Sorbonne Université, CNRS,
24, rue Lhomond, 
F-75005 Paris, France}
\affiliation{Universit\'e de Paris VII Denis Diderot, 5 rue Thomas Mann, 75013 Paris, France}

\email{zanchi@ens.fr}

%\date{\today }

\begin{abstract}
Effects of green tea tannin epigallocatechin-gallate (EGCG) on thermal-stress-induced amyloid fibril formation of reduced carboxymethylated bovine milk protein $\kappa$-casein (RCMK) were studied by dynamical light scattering (DLS) and small angle x-rays scattering (SAXS). Two populations of aggregates, micelles and fibrils,
dominated the time evolution of light scattering intensity and of effective hydrodynamic diameter. SAXS experiments allowed to resolve micelles and fibrils so that the time dependence of scattering profile revealed structural evolution of the two populations.   
The low-Q scattering intensity prior to an expected increase with time due to fibril growth, shows an intriguing rapid decrease which is interpreted as the release of monomers from micelles. 
This phenomenon, observed both in the absence and in the presence of EGCG, indicates that under thermal stress free native monomers are converted to amyloid-prone monomers that do not form micelles. The consumption of free native monomers results in a release of native monomers from micelles, because only native protein participate in micelle-monomer (quasi-)equilibrium.
This release is reversible, indicating also that native-to-amyloid-prone monomers conversion is reversible as well. We show that EGCG does not bind to protein in fibrils, neither does it affect/prevent the pro-amyloid conversion of monomers. EGCG hinders the addition of monomers to growing fibrils. These facts allowed us to propose kinetics model for EGCG-controlled amyloid aggregation of micellar proteins. Therein, we introduced the growth-rate inhibition function which quantitatively accounts for the effect of EGCG on the fibril growth at 
any degree of thermal stress.

\vspace{3mm}

\end{abstract}
%\pacs{....}

\maketitle

%\newpage

 \section{Introduction}

 Control and understanding of amyloid fibril formation \cite{Greenwald_2011} is emerging as a topic of importance in biomedical contexts (therapeutic applications for neurodegenerative diseases \cite{Bartolini_2010}, physiologically acceptable adhesion and tissue engineering \cite{Das_2018}), in functional materials \cite{Jones_2012} including nanotechnologies \cite{Wei:2017aa}, and in food and nutrition applications  \cite{Cao:2019aa}.

Here, we studied the perturbation/control of amyloid fibrils formation by addition of tanins. We focussed on a protein that forms micelles. This case is attracting considerable attention since early studies on amyloid beta (A$\beta$) peptide fibrillation   in which dynamic micellar oligomers determine the aggregation mechanisms \cite{Lomakin_1997,Morel_2018}. The micelle-fibril or oligomer-fibril interplay has been reported for a number of other proteins 
\cite{Redecke_2007,danino_2016,Orte2008,Vestergaard_2007}.
Thus, amyloid nucleus, hierarchical fibril formation, transient species including micelles/oligomers etc. have been identified in a number of cases by light scattering and time resolved small-angle x-ray (SAXS) scattering experiments \cite{Lomakin_2002,Vestergaard_2007,Oliveira_2009,Ortore_2011,Bolisetty_2011,Redecke_2007,pedersen_2010}. Most of these studies are based on multi-component analysis invoking singular-value-decomposition and other similar methods \cite{Herranz-Trillo:2017aa}, which is very efficient when the model oligomeric species are well-defined monodisperse objects (even when the system is composed of several different co-existing structurally different species) \cite{Schroer:2018aa}. These methods fail however in the presence of polydisperse oligomers or micelles.

Polyphenols and in particular small tannins show promising amyloid inhibitory activity for 
pathogenic proteins like $\alpha$-synuclein ($\alpha$-syn) and A$\beta$  \cite{Bartolini_2010,Jones_2012}. 
In nature, tannins protect plants from viruses, fungi, bacteria and higher herbivores by mechanisms based on physical interactions of native or oxidized  tannins with polysaccharides and proteins \cite{Pourcel_2007}. High affinity of tannins for proteins \cite{tannins_general, tannins_general_B}, makes tannins interesting for studying their impact on the formation of amyloid aggregates. A small tannin from the green tea EGCG is a particularly strong anti-amyloid agent. EGCG prevents protein from forming amyloid fibrils by co-assembling with proteins ($\alpha$-syn, peptide A$\beta$, $\kappa$-casein) into non-toxic spherical 20 nm protein-tannin aggregates 
\cite{Ehrnhoefer_2008,Hudson_2009,Lorenzen_2014}. This particularity of EGCG compared to other small polyphenols is assigned to its two gallate moieties \cite{Bieschke_2010} able to attach to two host sites on protein chains.
On the other hand, it is well known that the tannin-protein micellar assemblies are formed between natively non-structured proteins (salivary protein II1 and $\beta$-casein) and polyphenols\cite{Zanchi_2008jp,Zanchi_2008epl,Ma_2012}. Particularly strong tendency to integrate EGCG into micelles was reported in the case of milk casein \cite{Shukla:2009cr}.
A reversible fibrillar self-assembly of proteins with ability to up-take EGCG was also recently reported \cite{Hu:2018aa}. In the present work we focus on fibrillar aggregation of micellar $\kappa$-casein, a well-known protein from milk \cite{Holt:2013aa}, and investigate how the tannin epigallocatechin-gallate (EGCG,  458 g/mol) can be used to understand the aggregation mechanism and control the system. Reduced caboxymethylated $\kappa$-casein (RCMK) forms easily fibrils in vitro and micelle-fibril interplay is the key phenomenon in this respect
\cite{Farrell_2003,Thorn_2005,Leonil_2008,Ecroyd_2008,Ossowski_2012,danino_2016}. 
Clear inhibitory effect of EGCG on $\kappa$-casein fibril formation \cite{Hudson_2009,Carver_2010}, is that EGCG maintains $\kappa$-casein in its pre-fibrillar micellar state.

Two main events were evoked as rate-limiting (or critical) for converting micellar $\kappa$-casein (reversible, polydisperse)  into (quasi-)irreversible fibrils. In the first, proposed by the Rennes group \cite{Leonil_2008}, the conformational change within the micelle itself precedes the fibril formation from these oligomeric sub-blocks. This mechanism shares common features with the one proposed by Lomakin {\em et al.} for A$\beta$ system \cite{Lomakin_1997}. In both cases, the seeds for fibrillation are formed within micelles. However, in Rennes group approach the these "seeds" are in fact oligomeric building blocks for fibrils, while in Lomakin's approach the growth is governed by sequential (one-by-one) addition of monomers. On the other hand, according to the Adelaide group \cite{Ecroyd_2008}, the fibrillation is controlled by monomer release events from the micelles and the fibril growth is due to the accretion of fibrils from rapidly formed nuclei. Our analysis enables to establish a consensus interpretation that can reconciliate these two apparently mutually exclusive point of views. The resulting model fits all our experimental data.  Namely, we find strong indications of both processes: 1) monomer release from the micelles and 2) nuclei formation by conformational transformation of micelles. Furthermore, we find that resulting fibril accretion is strongly affected by EGCG. 

Fibril's growth  is most commonly studied by kinetics experiments for which the fibrillation mechanisms are represented by rate equations  and  the associated rate constants are found as fitting parameters \cite{Cohen_2011,Cohen:2011c,Grigolato:2019aa}.  The main difficulty in this regard is the polydispersity and tannin-dependent structure of micelles coexisting with fibrils all along the kinetics. We used DLS and SAXS to monitor kinetics and structure evolution of both micelles and fibrils upon thermal stress, and show how EGCG affects structures and kinetics. A simple kinetics model is proposed, based on reversible micelle-monomer exchange, native monomer to amyloid-prone monomer transformation, and subsequent growth. To implement the EGCG effect in the rate equation, a rate-inhibition function is constructed.

After presenting materials and methods in section \ref{S2}, the experimental findings are reported in section \ref{S3}. 
In section \ref{Model-free interpretations} details of kinetics and impact of EGCG are interpreted first within a model-free discussion, in which we assume no particular form of micelles nor theoretical kinetics model for aggregation. Most of our conclusions are validated already at this level. In section \ref{Modelling}, with the help of structural information from fitted SAXS data,
kinetics model based on rate-equations is constructed. 
In section \ref{Discussion} we discuss our interpretations in regard to other possible mechanisms of fibril growth and of mechanisms by which the EGCG impacts the system. Conclusions are given in section \ref{Conclusions}.

\section{Materials and Methods}
\label{S2}

 \subsection{Buffer}
 All experiments were carried out in a 50 mM phosphate buffer at pH=7.2 obtained by mixing 50 mM solution of NaH$_2$PO$_4$ with 50 mM solution of NaHPO$_4$ at a volume fraction 1.88:1. Chemicals were purchased from Sigma. 
  
 \subsection{Preparation of the reduced carboxymethylated $\kappa$-casein (RCMK)}
 
 The $\kappa$-casein was purchased from Sigma.
 The reduced caboxymethylated derivative of $\kappa$-casein (denoted as RCMK) was prepared
by reduction of the $\kappa$-casein disulfide linkages and subsequent carboxymethylation as described in literature
\cite{Schechter_1973}. Resulting RCMK was dialysed in 50 mM phosphate buffer at pH 7.2 and then freeze-dried to get amorphous powder. The stock solutions were recovered from the powder by adding the same volume of Milli-Q water as the RCMK solution contained before the freeze-drying. 

\subsection{Light scattering (LS) experiments}
\label{LSsub}
  
Light-scattering experiments were performed on a BI-200SM
goniometer (Brookhaven instruments) equipped with two photomultipliers and cross-correlator BI-9000AT, a Mini-L30 (30 mW, $\lambda _0=637$ nm) compact diode laser, and a circulating temperature-controlled water bath (PolyScience, USA). 
Samples were prepared from filtered (100 nm PVDF filter) stock solutions of protein and of tannin.  After mixing, the 500 $\mu $L samples were re-filtered  and readily subjected to in situ evolution at 45 $^o$C. Scattering intensity at 90 $^o$ was monitored. The analysis of the homodyne intensity--intensity correlation function
$G(q,t)$  was performed using the cumulant method \cite{mailer2015particle}.  
Accordingly, the mean sphere-equivalent (Stokes) hydrodynamic radius $R_{H{\mbox{\tiny eff}}}$ is obtained from the mean relaxation rate $\Gamma$ of $G(q,t)$. 
\begin{equation}
\Gamma =\frac{k_B T}{6 \pi \eta R_{H{\mbox{\tiny eff}}}} q^2\; ,
\end{equation}
where $\eta$ is water viscosity and $q=\frac{4 \pi}{\lambda} \sin (\theta /2)$ is the scattering vector for scattering angle $\theta$.

In the present case the system is composed of micelles and fibrils, both polydisperse and with comparable diffusion times. Consequently, any attempt to resolve the light-scattering correlation function in terms of 2 populations would fail. Therefore, we adopted a well-established Lomakin's approach \cite{Lomakin_1997} to micelle-fibril system. 
For a polydisperse and multi-population system (for us: micelles and fibrils) the scattering intensity is given by :
\begin{equation}
I-I_b=  K_{\mbox{\tiny mic}} \sum _i i^2 f_{\mbox{\tiny mic}}(i) m_i + K_{\mbox{\tiny fib}} \sum _i i^2 f_{\mbox{\tiny fib}}(i) n_i
\label{I}
\end{equation}
and the effective hydrodynamic radius from cumulant is obtained from
\begin{widetext}
\begin{equation}
\frac{1}{R_{H{\mbox{\tiny eff}}}}=  \frac{K_{\mbox{\tiny mic}} \sum _i i^2 f_{\mbox{\tiny mic}}(i) m_i/R_{H,{\mbox{\tiny mic}}}(i) + K_{\mbox{\tiny fib}} \sum _i i^2 f_{\mbox{\tiny fib}}(i) n_i /R_{H,{\mbox{\tiny fib}}}(i)}{K_{\mbox{\tiny mic}} \sum _i i^2 f_{\mbox{\tiny mic}}(i) m_i + K_{\mbox{\tiny fib}} \sum _i i^2 f_{\mbox{\tiny fib}}(i) n_i}\; ,
\label{R}
\end{equation}
\end{widetext}
where $I_b$ is time independent background intensity, $K_{\mbox{\tiny mic}}$ and $K_{\mbox{\tiny fib}}$ are instrument-dependent constants, proportional to refraction index gradient for micelles and for fibrils, $f_{\mbox{\tiny mic}}(i)$ and $f_{\mbox{\tiny fib}}(i)$ being the form factors for light scattering from micelles and from fibrils 
i-mers, $m_i$ and $n_i$ being the number densities of micelles and fibrils i-mers, while $R_{H,{\mbox{\tiny mic}}}(i)$ and 
$R_{H,{\mbox{\tiny fib}}}(i)$ stand for effective Stokes radii for micelles and for fibrils i-mers. 

We simplified these equations by applying the following approximations.
First, we know that micelle gyration radius (less than $20$ nm) is smaller than $1/q=\lambda _0/11.8=54 $ nm, $q$ being the scattering vector in water at 90$^o$ scattering angle and $\lambda _0=637$ nm. For this reason we put $f_{\mbox{\tiny mic}}(i)=1$. 
Second, we assumed that micelles and fibrils scattering contrast was the same: 
$K_{\mbox{\tiny mic}}=K_{\mbox{\tiny fib}}=K$.
Strictly speaking, this is not {\em a priori} the case because, as we will show later, tannin is not tolerated in fibrils but is present in micelles. However, the error is negligible because protein and EGCG have, surprisingly, close LS contrasts.  
Third, the hydrodynamic radius of micelles is independent on $i$ for a large majority of micelles, so we put $R_{H,{\mbox{\tiny mic}}}(i)=\mbox{cte}.(i)\equiv R_{H,{\mbox{\tiny mic}}}$.
On the other hand, variation of the hydrodynamic radius of fibrils with fibril length $L$, $R_{H,{\mbox{\tiny fib}}}(L)$ was typically obtained from a calibration curve, shown in the "Results" section, while $L$ and $i$ are related by $L=i/\lambda$, where $\lambda$ is the number of protein monomers per nanometer of fibril length. 
Finally, the fibrils form factor $f_{\mbox{\tiny fib}}(i)$ was approximated by an infinitely thin rod form factor \cite{Pedersen_Lindner_Zemb}, because the width of the fibrils ($\approx 30$nm) is inferior to $1/q$.
\begin{equation}
f_{\mbox{\tiny fib}}(i)\approx f_{\mbox{\tiny rod}}\hspace{-0.5mm}(qL(i)) \; , \hspace{1.5mm}f_{\mbox{\tiny rod}}\hspace{-0.5mm}(x)\equiv \frac{2 \text {Si} (x)} {x}-4 \frac{\sin^2(x/2)}{x^2}\; .
\end{equation}
In order to relate the measured scattering intensity and effective hydrodynamic radii to our simple kinetics model, only first two moments of  the distributions $m(i)$ and $n(i)$ are considered. We use: $M=\sum _i i  \, m_i$ (concentration of monomers associated in micelles), $\bar{m}=\sum _i i \,  m_i / \sum _i   m_i $ (average micellization number), $N_0=\sum _i   n_i $ (concentration of fibrils) and $N_1=\sum _i   i\, n_i $ (concentration of  monomers associated on fibrils). Consequently, the average fibril length is $L=N_1/(N_0 \lambda)$.
Altogether, allowing $M$, $N_0$ and $N_1$ to evolve in time during fibrillation, the expressions (\ref{I}) and (\ref{R}) reduce to
\begin{equation}
I(t)-I_b=  K \left[ \bar{m} M(t) + f_{\mbox{\tiny rod}}\hspace{-1mm}\left(\frac{11.8 L(t)}{\lambda _0 }\right) \frac{N_1^2(t)}{N_0(t)}\right] \; ,
\label{Isimp}
\end{equation}
\begin{widetext}
\begin{equation}
\frac{1}{R_{H{\mbox{\tiny eff}}}(t)}=  \frac{\bar{m} M(t) R_{H,{\mbox{\tiny mic}}}^{-1}+ f_{\mbox{\tiny rod}}\hspace{-1mm}\left(\frac{11.8 L(t)}{\lambda _0 }\right) \frac{N_1^2(t)}
{N_0(t)} R_{H,{\mbox{\tiny fib}}}^{-1}\hspace{-1mm}( L(t) ) }
{\bar{m} M(t) + f_{\mbox{\tiny rod}}\hspace{-1mm}\left(\frac{11.8 L(t)}{\lambda _0}\right) \frac{N_1^2(t)}{N_0(t)}}
\label{Rsimp}
\end{equation}
\end{widetext}

In order to interpret correctly the DLS results, we checked that the scattering intensity was almost entirely composed of the scattering by protein-protein and by protein-tannin large assemblies, whose weight was at least of the order of a micelle's mass ($\sim$ 0.5 MDa).
The refractive index for tannin and protein is approximately the same, giving, for our setup $(\theta = \pi/2$, pinhole 200 microns), for a single population, the average scattering intensity of the form $I={\cal K}C\cal{M}$, with ${\cal K}=0.1\pm 0.01$ kcps L (g Da)$^{-1}$,
$C$ being the concentration in g/L and ${\cal M}$ the molar mass. Accordingly, the scattering intensity from dissolved  EGCG is $\sim$200 times lower than the scattering from protein micelles at same concentration by weight.  The contribution to the scattering intensity of monomeric protein is also negligible: the micelles are 30 times heavier and the fraction of monomers is of the order of $\sim 10 \%$ of the total protein amount. We also verified that no tannin self-association took place, by comparing LS from the samples with and without tannins before incubation.
Furthermore, we observed no self aggregation of tannin induced by temperature used for incubations. Modifications of the scattering intensity due to absorption of the light by tannins were also negligible. 

Adding tannins at the concentrations used in this work did not significantly  change the physicochemical properties of the solvent (pH, conductivity, viscosity, water activity). Therefore, the physical quantities measured by DLS were predominantly reflecting protein-protein and tanin-protein interactions.

\subsection{Small angle X-ray scattering (SAXS) experiments} 
 
The synchrotron radiation X-ray scattering measurements were performed at the SWING beam line of the Soleil synchrotron facility in Saclay, France.
The incident beam energy was 12 keV. In most experiments the sample to detector (Aviex CCD) distance was set to 1817 mm, covering the $Q$ range from 0.06 to 7 nm$^{-1}$. Samples of 30-50 $\mu L$ were injected in the capillary by a thermostated flow-trough device between two air bubbles and the flow was continuous during the SAXS data acquisition, in order to avoid sample degradation by the X rays. 
Static experiments were temperature-controlled at 25 $^{\circ}$C. For the kinetics experiments the temperature of samples and of the flow-through capillary was set at 40 $^{\circ}$C. 
Typically 40 successive frames
 of 0.5 s each were recorded for both sample and pure solvent. Each frame was first angularly averaged and the final spectrum and experimental error were obtained by averaging over all frames and subtracting the pure solvent spectrum from the sample spectrum. Intensities were scaled to absolute units using the scattering of water.

\subsubsection{Structural model for micelles}
\label{Structural}
In order to capture the structure of micelles, we used the Svaneborg-Pedersen model for core-crown micelles with partially rigid crown polymers \cite{pedersen00}.
In this model it is supposed that the micelles consist of a core with the radius $R_0$ to which are attached the polymers, see  (\ref{micelle_ellCylShell}). The conformations of polymers in the crown are not a simple gaussian coils because the polymers interact with each other via excluded volume or by electrostatic repulsion. The corresponding stretching of chains is included in the model by assuming that the chains are partially rigid over the length $l$ starting from their end attached to the core. The remaining part of the chain is in the gaussian random walk conformation.
In the case of simple diblock copolymers with a hydrophilic block $\alpha$ and hydrophobic block $\rho=1-\alpha$, all hydrophilic parts of polymers constitute the crown, and the remaining, hydrophobic parts, are in the core.  Assuming that all residue have the same scattering contrast the Svaneborg-Pedersen form factor for micelles with the aggregation number $m$ (assumed to be equal to the number of polymer segments in corona) writes
\begin{widetext}
\begin{equation}
%\begin{eqnarray}
P(q)=F_{\mbox{mic}}^{R_0,m,R_{gc},l}(q)=\rho^2F_1^2(qR_0)+\alpha^2\frac{F_c(q)}{m}+
\alpha^2\frac{m-1}{m}S_{cc}(q)+2\alpha \rho S_{cs}(q)F_1(qR_0) \; .
%\end{eqnarray}
\label{SP}
\end{equation}
\end{widetext}
The first term comes from the scattering from the spherical core alone:
$F_1(q)=3 (\sin (q R_0) - qR_0 \cos (q R_0))/(qR_0)^3$. The term $F_c(q)$ comes from individual chains in the crown and depends on $R_{gc}$ and $l$, being respectively
the gyration radius of the gaussian part of the chain and the length of the rod part.
The term $S_{cc}(q)$ is the inter-chain interference contribution and  $S_{cs}(q)$ contains core-crown interference. The analytical forms of functions $F_c(q)$,  $S_{cc}(q)$ and $S_{cs}(q)$ are given in \cite{pedersen00}.
In this work, the quantity of tannins bound to proteins in micelles considerably contributed to scattering intensity.
Therefore we modified the model (\ref{SP}) by adding supplementary contrast into the core of micelles, 
by assuming 
\begin{equation}
\rho =1-\alpha +\phi \; ,
\label{phi}
\end{equation}
 where $\phi$ is the increment of the density of the micelle core per protein. This increment comes {\em a priori} from both tannins sticked to proteins and supplementary protein brought by tannin into the core.
We included polydispersity in our analysis by assuming that the scattering intensity is given by:
\begin{equation}
P_{\sigma}(q) =\frac{1}{\sigma \sqrt{2 \pi}}  \int_0^\infty d R \; {e^{-\left[\frac{R - R_0}{\sigma/2}\right]^2}} F_{\mbox{mic}}^{R,m,R_{gc},l}(q)\; ,
\label{polydisp}
\end{equation}
where $\sigma$ is a log-normal distribution width. In that way oscillations due to spherical micelle core were averaged out. Polydispersity in other parameters: $N$, $R_{gc}$ and $l$ does not change substantially the scattering curves. Therefore we used simply their average values to fit the data.

%%mod:
The resulting expression for scattered intensity of protein-tannin micelles is 
\begin{equation}
I(q) = c_1 K_p m {\cal M}_0 P_{\sigma}(q)\; ,
\label{scatintens}
\end{equation}
where $c_1$ is protein concentration in g/L, ${\cal M}_0=19000$ g/mol is the molecular weight of $\kappa$-casein, $m$ is the average aggregation number and $K _p=N_A(V_p \Delta \rho_p /{\cal M}_0)^2$ is the SAXS scattering constant for protein, expressed over the volume of solvated protein $V_p$, corresponding excess scattering length density $\Delta \rho_p$, and molecular weight ${\cal M}_0$. For the X-ray contrast of the protein we used $\Delta \rho_p=0.35 \rho _{H_2O}$  and of the tannin we had $\Delta \rho_t=0.39 \rho _{H_2O}$, where $\rho _{H_2O}=9.46\times 10^{10}$ cm$^{-2}$ is the SAXS scattering length density for water \cite{Schurtenberger_Lindner_Zemb}.
$\Delta \rho_t$ was determined from the absolute forward intensity $I_t$ of monodispersed tannin dimer with (molecular weight $M_t=2 \times 288 +2$ g/mol) at concentration 
$t=5$ g/L, using the relation
\begin{equation}
\Delta \rho_t= d_t \sqrt{\frac{N_A I_t}{t M_t}}     \; ,
\end{equation}
where tannin relative density  $d_t$=1.55 was determined by pycnometer. Excess electronic densities corresponding to above contrasts were 118 electrons per nm$^3$ for protein and 132 electrons per nm$^3$ for tannin. 

From our fits we get the parameter $\phi$. We cannot distinguish between tannins' and proteins' contributions in $\phi$: a detailed composition of the core in terms of tannins and proteins is not reachable by SAXS. 
However, from the quantity $\phi$, the evolution of $m$ and $R_0$ we can find the excess electron density in the core 
\begin{equation}
n_e =(1+\frac{\phi}{\rho})\frac{m}{m_0}\left(\frac{R_{00}}{R_0}\right)^3 n_{e0}\; , 
\label{exdens}
\end{equation}
where $m_0$, $R_{00}$ and $n_{e0}$  are respectively the aggregation number, the core radius and the excess electron density in the core for micelles without tannin.

Alternative model for $\kappa$-casein micelles, proposed by Lund group \cite{Ossowski_2012}, assumes that the core has oblate ellipsoidal form. In this case, the form factor writes similar to (\ref{SP}), where the sphere term of $F_1^2(q)$ is replaced
the oblate spheroid form factor $F_{\mbox{\tiny ells}}(q)$, given by \cite{Lorenzen_2014} :
\begin{equation}
F_{\mbox{\tiny ell}}(q)=\int_0^{\frac{\pi}{2}} \left[F_1\left (q R\sqrt {\epsilon^2\cos^2 (\alpha) + \sin^2 (\alpha)}\right) \right]^2 
\sin \alpha \; d  \alpha \; ,
\label{elli}
\end{equation}
$R$ being the short spheroid radius and $\epsilon$ the long-to-short radii ratio.

In our analysis both spherical and oblate models were tested and the choice of better model was the spherical one, although the differences were small. Dependence of the micelles structure and polydispersity on the preparation protocol influences the form factor by an amount comparable to the difference between spherical and oblate models, see inset of fig.\ref{f5}. For these reasons we stick to the spherical polydisperse model, keeping in mind that all our conclusions would be qualitatively the same if we were to use the oblate model.

\subsubsection{Structural models for fibrils}
 
\begin{figure*}[t]
  \centering
  \includegraphics[width=14cm]{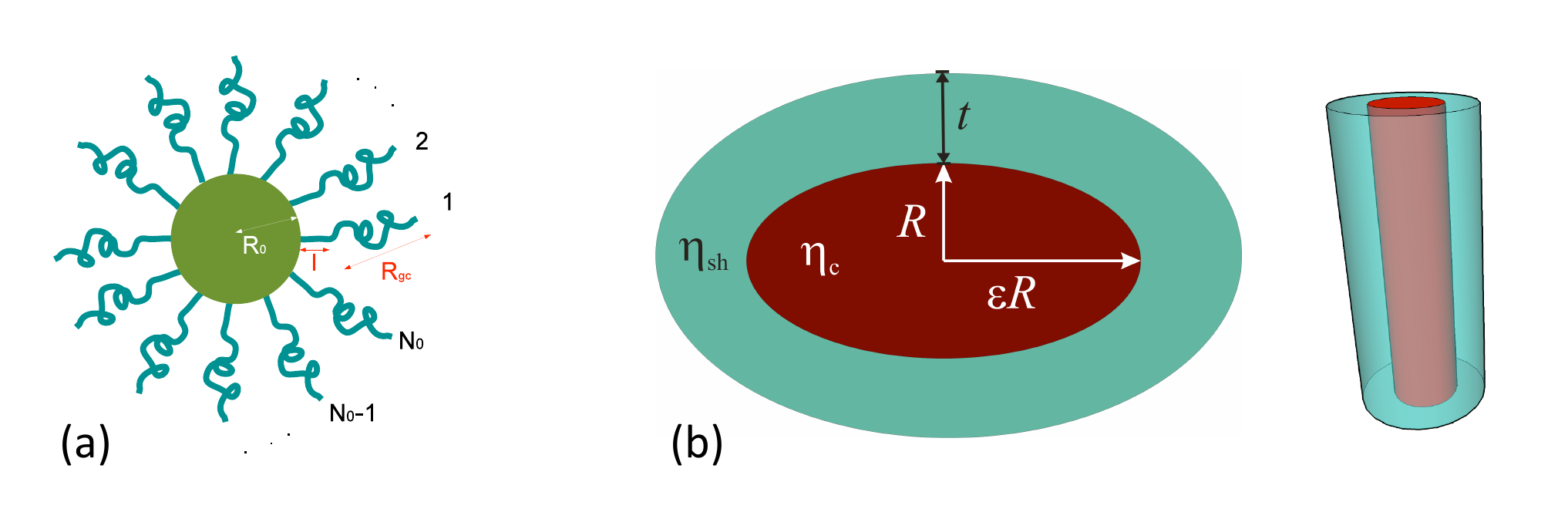}
  \caption{Structural models used to calculate the form factor for fitting SAXS data of (a) the protein-tannin micelles and of (b) the amyloid fibrils.}
  \label{micelle_ellCylShell}
\end{figure*}

 To fit the SAXS profiles of the fibrils, we used two models: (i) long homogenous cylinder and (ii) long core-shell cylinder with elliptical cross-section, implemented in the package for fitting  small angle scattering data SASFIT \cite{SASFIT}. For calculating the form factor for long homogenous cylinder we used the routine LongCylinder, implying the circular cylinder radius $R$ and the cylinder length $L$ as fitting parameters. The routine uses the Porod approximative formula for long cylinder, valid for $L > 2R$ \cite{porod_1982}. 
The form factor of a core-shell cylinder with elliptical cross-section was calculated using the routine ellCylShell1, according to \cite{SASFIT}.  The model depends on the following fitting parameters: the core short semi-axis $R$, the eccentricity $\epsilon$ of cross-section (the long semi axis is $\epsilon R$), the cylinder length $L$,  the shell thickness $t$, the scattering length density for the core $\eta_c$ and the scattering length density for the shell $\eta_s$. The model is shown on fig.\ref{micelle_ellCylShell}.

%*********************************************************************  
%*********************************************************************

  \subsection{Atomic force microscopy (AFM)}
 
Substrates for imaging were prepared as follows: an aliquot of the sample was cooled down to ambient temperature. Then a freshly cleaved mica plate was dipped in 10 times diluted sample for 30 s, and rinsed by deionized water for 30 s. The mica plate was finally dried at 40 $^{\circ}$C for 15 min in an oven.
 Atomic Force Microscopy (AFM) measurements were performed using a 5100 Atomic Force
Microscope (Agilent technologies - Molecular Imaging) operated in a dynamic tip
deflection mode (Acoustic Alternating Current mode, AAC). All AFM experiments were
done using Silicon Probes (Applied NanoStructures-FORT) in the tapping mode
with spring constant 3 N/m at 73 kHz, close to the cantilever's resonant frequency, with a resolution of 512$\times$512 pixels.

 \section{Results}
 \label{S3}

\subsection{Fibril growth kinetics with a range of EGCG concentrations: DLS data}

We used light scattering measurements to asses quantitatively the kinetics of fibril growth and the effect of the EGCG tannin in solutions of $\kappa$-casein. In non-reducing conditions, it is important to avoid the formation of non-amyloid large aggregates due to disulfide bonds (over 1 MDa at 25 $^{\circ}$C and pH 6.5-7.5) \cite{Leonil_2008}. This can be achieved by capping cysteines in $\kappa$-casein by caboxymethylation \cite{Farrell_2003}. At 25 $^{\circ}$C and pH 6.5-7.5 caboxymethylated $\kappa$-casein (RCMK) forms core-crown micelles composed of approximately 30 proteins, with the critical micellization concentration (CMC) of approximately 0.05 $\%$.\cite{Ecroyd_2010} The micelles have the core radius of 6-7 nm and the crown segments reach up to the radius of 14.7 nm \cite{deKruif_1991,Ossowski_2012}. Solutions of 3 g/L$^{-1}$ RCMK in 50 mM phosphate buffer pH 7.2, with varying EGCG concentrations, were stable at room temperature. The aggregation kinetics was triggered by inserting the sample in the scintillation vial at 45 $^{\circ}$C (cf. Methods section).
At 45 $^{\circ}$C the fibrils grew considerably faster than at 37 $^{\circ}$C, and the inhibition by EGCG was lower: while at 37 $^{\circ}$C an 
equimolar concentration of EGCG to RCMK was sufficient to inhibit completely the fibril growth\cite{Hudson_2009}, at 45 $^{\circ}$C we needed 15 EGCG-s per protein to get the same effect.

\begin{figure*}[t]
  \centering
  \includegraphics[width=15cm]{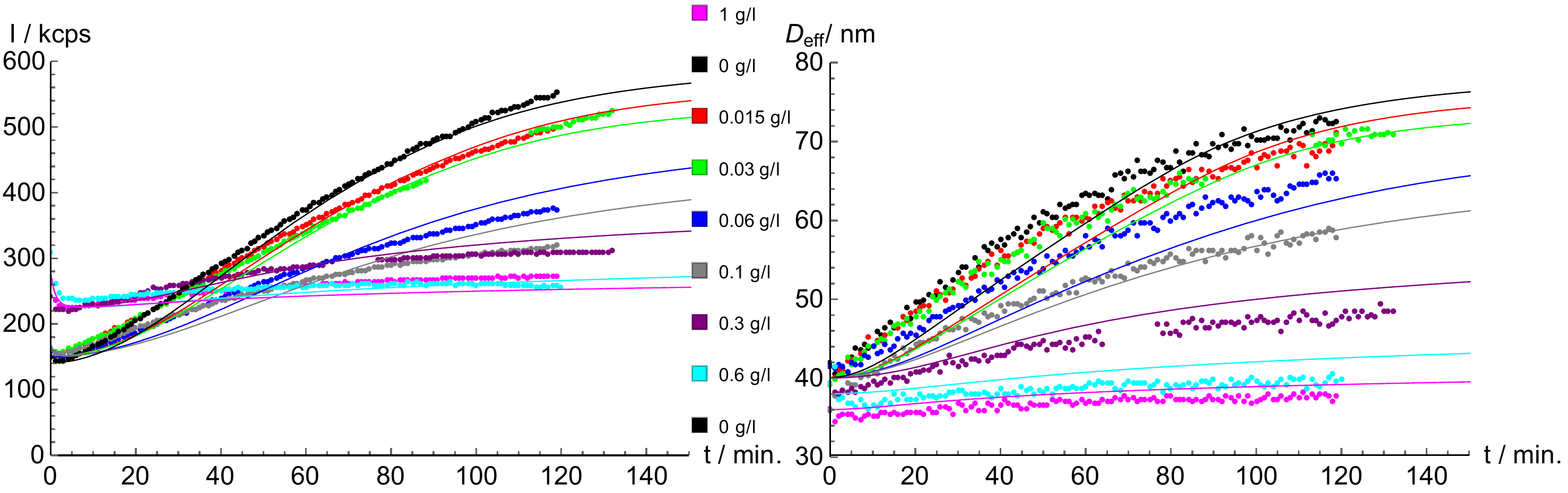}
  \caption{Evolutions of scattering intensities $I(t)$ and of effective hydrodynamic radii $D_{\text{\tiny eff}}(t)$ during 120 min of RCMK alone and in presence of a range of EGCG concentrations. Solid thin lines are the fits to our model.
}
  \label{f2}
\end{figure*}

We focus on two characteristic quantities: the scattering intensity and the effective hydrodynamic diameter, keeping in mind that both observables originate from both micelles and fibrils.
Evolutions of the scattering intensities and of the effective hydrodynamic radii $D_{\text{\tiny eff}}=2R_{H{\text{\tiny eff}}}$ during 120 min of RCMK alone and in the presence of a range of EGCG concentrations are shown in Fig.\ref{f2}. 
Full thin lines are the fits to our model described in the next section.

 At the room temperature of 25 $^{\circ}$C (data at t=0) all samples contained oligomeric objects, presumably micelles, with $D_{\text{\tiny eff}}$ between 35 and 40 nm. The scattering intensity at room temperature increased with the addition of EGCG. On the other hand, the hydrodynamic radii at room temperature decreased upon addition of EGCG. 

The scattering intensity of all samples first decreased with increasing incubation time, reaching a minimum after typically $\sim4$ min. We notice that the magnitude of intensity drop increased with increasing amount of tannin.
At times longer than ca. 4 min, the increase of $I(t)$ indicated the growth of fibrils. The evolution $D_{\text{\tiny eff}}(t)$ was an initial increase and final plateau as well. Both the growth rate and the plateau decreased with increasing the EGCG concentration. A tannin-dependent final plateau suggests that EGCG blocks the fibrillation rather than just slowing it down.

%*********************************************************************  
%*********************************************************************

\begin{figure}[t]
  \centering
  \includegraphics[width=9cm]{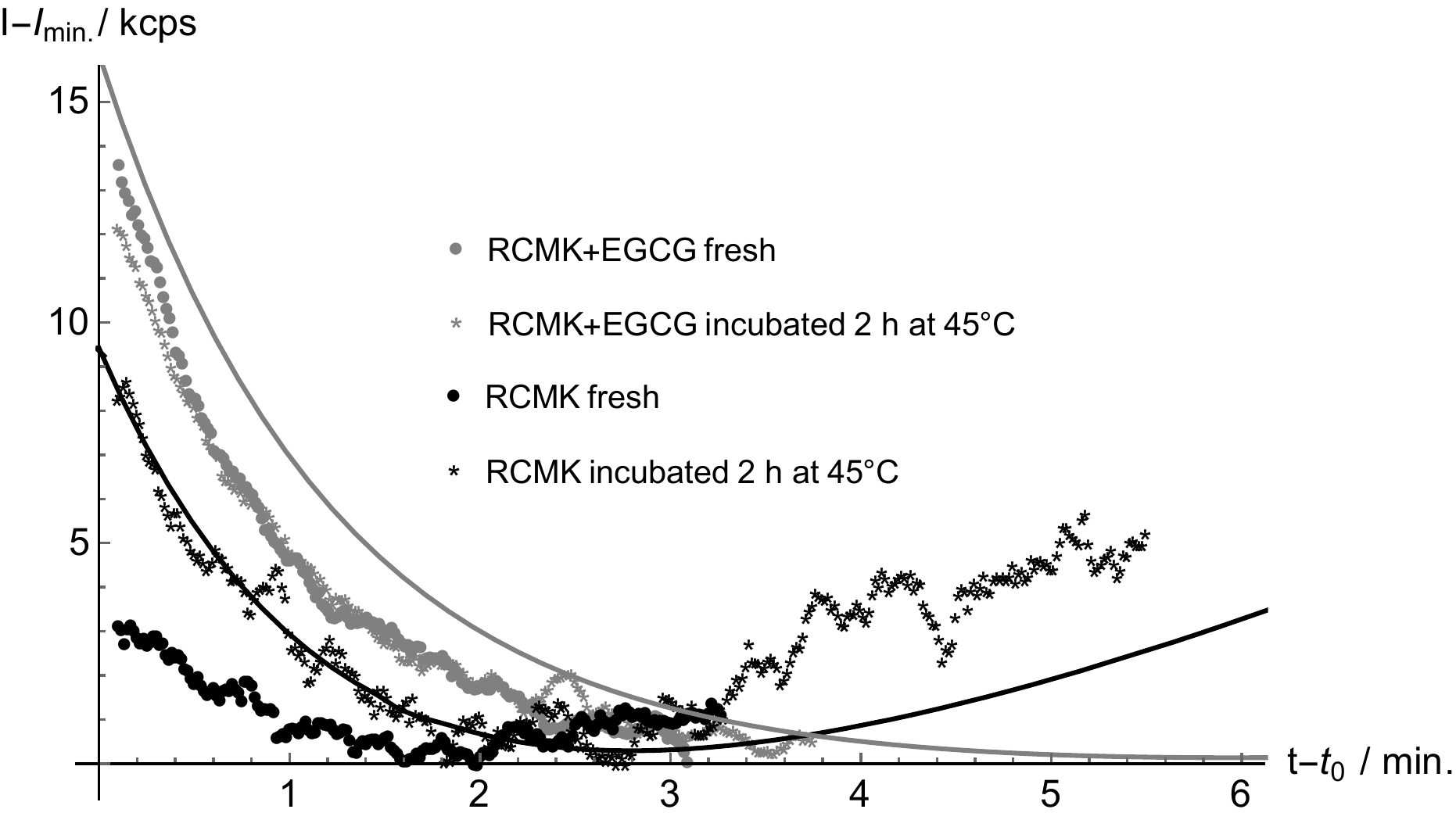}
  \caption{Reversible release and uptake of monomers by micelles: Light scattering intensity profile in the beginning of kinetics compared to the kinetics from 2h incubated, cooled and re-heated samples for RCMK (3g/L) alone and RCMK with 1 g/L of EGCG. Solid lines are results of our kinetics model: black line for RCMK alone and gray line for RCMK with 1 g/L of EGCG.}
  \label{f3}
\end{figure}

We first comment in this paragraph the intriguing initial decrease of the scattering intensity. 
Interestingly, an initial decrease, although slower, was detected in thyoflavin fluorescence assay reported in \cite{Hudson_2009}. Authors interpreted it as a transient effect due to temperature stabilisation. 
In our DLS experiment, just as in the ones of the ref. \cite{Hudson_2009}, no change was detected in hydrodynamic radius $D_{\text{\tiny eff}}$. However, the larger magnitude of variation observed in the present scattering intensity suggests a contribution of some reorganization of micelles (ex. fragmentation).
We hypothesized that the initial decrease in scattering may be due either to the release of small species from micelles or some re-organization between micelles, which has to remain essentially of the same size but can nevertheless change the average aggregation number $\bar{m}$. 
Reversibility of this initial variation was assessed as a criterion to distinguish equilibrium assemblies of native-like monomers, from other likely irreversible reorganizations involving fibrils. Samples were incubated for 2h at 45$^{\circ}$C, and cooled down to 25$^{\circ}$C in a separate thermostated bath (for typically 10 minutes), and then re-inserted the vial back to 45$^{\circ}$C in the DLS device. The variation of scattering intensity collected from these re-heated samples, are compared in fig.\ref{f3} to the intensity profile recorded in the beginning of kinetics for two extremal cases: RCMK (3g/L) alone and RCMK with 1 g/L of EGCG. In the case of high EGCG (blocked fibrillation), $I(t)$ reached a quasi plateau after a few minutes of decrease, which may reflect a new equilibrium between micelles and monomers. During the second heating step, variation of intensity was similar to the one obtained in the first heating step, suggesting that the scattering reflect a reversible, temperature-dependant equilibrium. In the case of RCMK alone, a considerable amount of protein has been converted to fibrils when the second heating was implemented (see below). During the second heating step, after decreasing over a characteristic time of 2-4 minutes, the intensity increased with a growth rate that matches the kinetic profile of a sample whose kinetics was not arrested by cooling. This indicates that cooling and re-heating did not change the number of fibrils and did not produce any new seeds. But without EGCG, the magnitude of the initial decrease in scattered intensity was different during the first and second heating steps, suggesting that the presence of fibrils promotes a temperature-sensitive conversion.

Remarkable fact is that micelles dissolve with increasing temperature, which is the opposite of what one expects from standard models based on the amphiphilic nature of protein. 
In particular, the effect of increasing temperature on $\beta$-casein, a milk protein not prone to form fibrils, is an increase of the scattering intensity (data not shown), reflecting the "usual" behaviour, i.e. conformity with the standard model \cite{Mikheeva_2003}.
As we will discuss later, this abnormal temperature dependence of RCMK micelles is the key point for understanding fibrillation and the effects of tannin on it.

%*********************************************************************  
%*********************************************************************

\subsection{Structural data: SAXS and AFM experiments}

The main difficulty in studying micelle-fibril system by DLS is that it is impossible to resolve two polydisperse species with similar relaxation times. Furthermore, based on the results above, one has to include a contribution from micelles dissolution (or reorganisation) triggered by the shift in T. The effect is enhanced by the addition of EGCG.
Using SAXS experiments enables to (i) get insight into the evolution of both micelles and fibrils upon thermal stress and (ii) decipher 
the modifications in micelle population occurring upon sudden temperature increase and to relate this effect to fibril growth mechanisms.    

\begin{figure}[t]
  \centering
  \includegraphics[width=9cm]{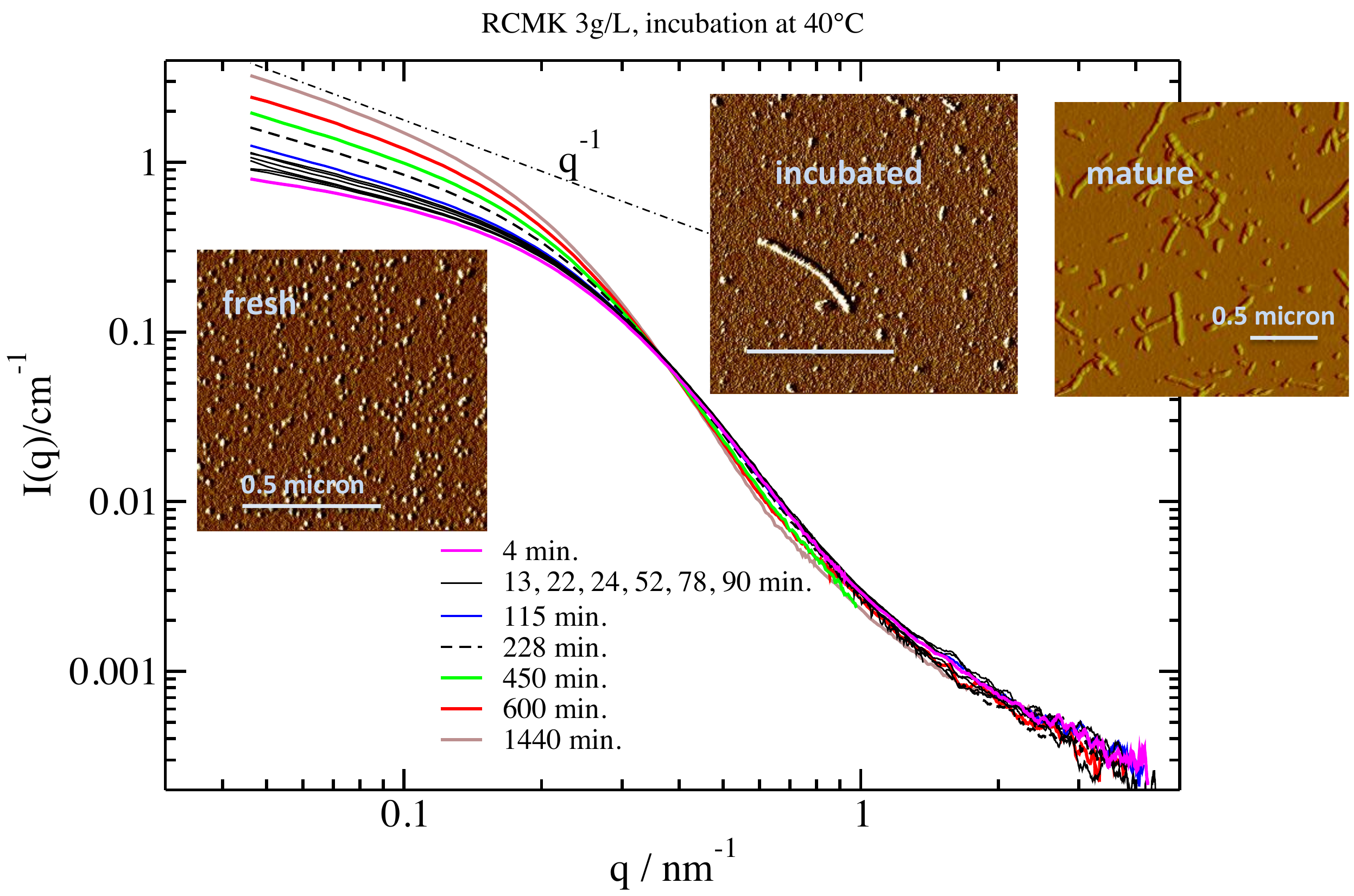}
  \caption{Time series of SAXS profiles for RCMK 3 g/L incubated at 40 $^{\circ}$C.  
  AFM scans of fresh sample, one day incubated sample, and mature sample (3 days incubated) dried at mica plates are show in insets. 
The average fibril length in mature samples was $L_{\infty}=450\pm 70$nm.}
  \label{f4}
\end{figure}

Time evolution of the SAXS profile for 3 g/L RCMK incubated at  40 $^{\circ}$C for 24 hours is shown in fig.\ref{f4}. 
It was not possible to incubate at higher temperature because of instrument limitations. Consequently, typical fibril kinetics are at the scales of days and not of hours, in agreement with data in literature.\cite{Ecroyd_2010} To illustrate, AFM scans of fresh and incubated samples dried at mica plates are show in insets of fig. \ref{f4}. Both micelles (small dots) and fibrils are clearly visible. A typical scan of a final, mature sample, where all protein is converted to fibrils is also shown. 
The average fibril length, calculated over several scans, was $L_{\infty}=450\pm 70$ nm.

The first SAXS profile of fig. \ref{f4} corresponds to 4 minutes incubation. At times shorter than 4 minutes, the low-Q scattering intensity decreased, which was consistent with results from light scattering. At times beyond 4 minutes, the low-$Q$ part of the SAXS profiles increased and tended toward a $Q^{-1}$ scaling, indicating that the protein progressively formed rod-like aggregates. Another interesting point in fig. \ref{f4} is that all scattering profiles have a common intersection, or an isobestic point, which is an indication that this set of profiles may come from a sum of contributions of two populations whose relative fraction change with time. Of note, the contribution from individual (non assembled) protein monomers is negligible in present conditions, as is negligible (for the same reasons of low molar mass) the contribution from unbound individual tannin molecules.

\begin{figure}[t]
  \centering
  \includegraphics[width=9cm]{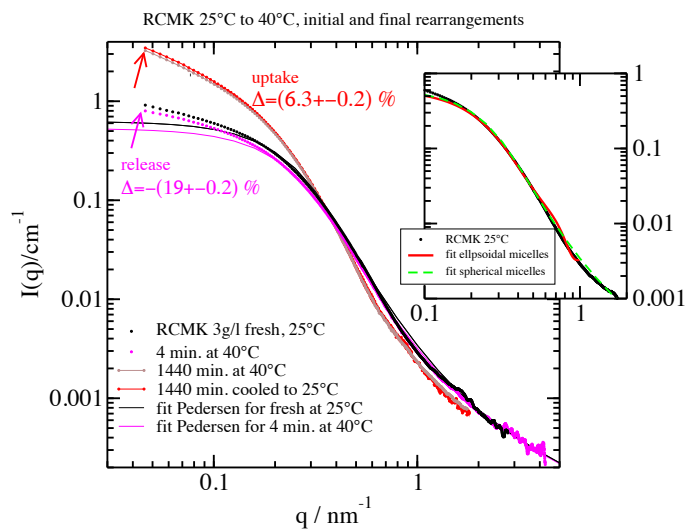}
  \caption{SAXS profiles for RCMK 3 g/L : initial drop of forward intensity upon T-jump from 25 to 40 $^{\circ}$C and uptake upon cooling back to 25 $^{\circ}$C of the sample after 1440 minutes of incubation. Inset shows the experimental profile fitted by both spherical (\ref{SP}) and ellipsoidal micelles form factors (\ref{elli}).}
  \label{f5}
\end{figure}

During the first minutes of incubation the low-Q scattering intensity decreased, confirming our findings from the light scattering experiment. 
Fig.\ref{f5} shows the SAXS profiles  just before and 4 minutes after T-jump from 25 to 40 $^{\circ}$C in the beginning of kinetics. The profiles are  compared to the final one (incubation for 1440 min.) and to the one obtained after cooling back to 25 $^{\circ}$C the same sample.
The initial decrease of the forward intensity was 0.095$\pm 0.01$ cm$^{-1}$, while the final increase upon re-cooling was 0.21$\pm 0.02$ cm$^{-1}$. This result confirms findings from DLS experiment: T-dependant variation in the fraction of small species is higher when some fibrils are present.

\begin{figure}[t]
  \centering
  \includegraphics[width=9cm]{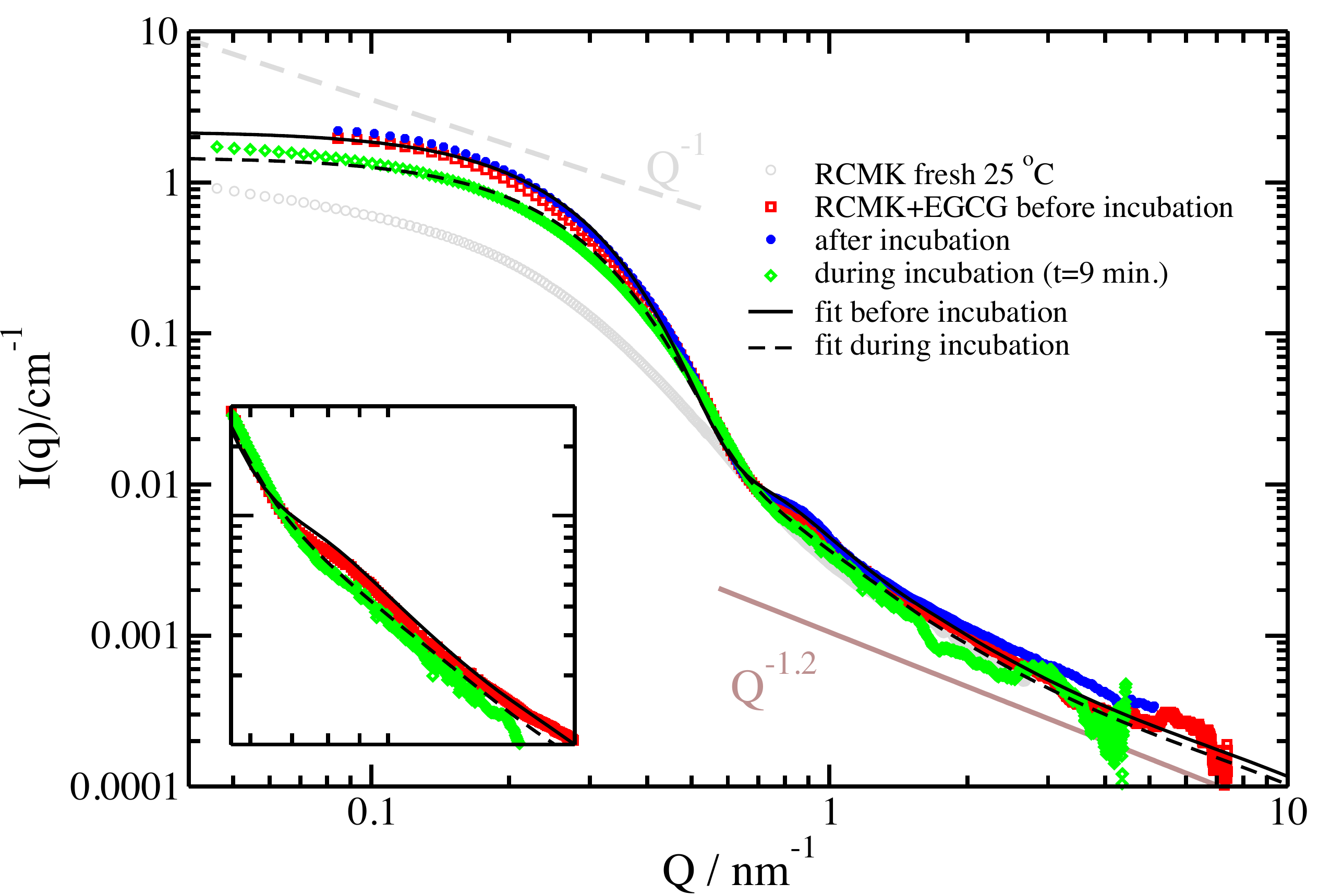}
  \caption{SAXS spectra of RCMK with EGCG: before incubation, during incubation at 40 $^{\circ}$C and after incubation, re-cooled to 25 $^{\circ}$C.}
  \label{f8}
\end{figure}

To explore the structure of the tannin-protein complex when fibril growth is essentially blocked,
a sample composed of RCMK 3 g/L and EGCG 1 g/L was studied by SAXS.
 Fig.\ref{f8} shows the SAXS curves before incubation, after 9 minutes of incubation at 40 $^{\circ}$C and after 70 min. of incubation, cooled down to 25 $^{\circ}$C. The fresh sample (before incubation) had a well defined Guinier regime, in contrast to RCMK alone, which had an upturn at low Q, due to some aggregation. This indicates that EGCG  stabilises micelles. This effect has already been reported in \cite{Zanchi_2008epl} for the case of $\beta$-casein micelles. 
As soon as the sample was heated to 40 $^{\circ}$C, the forward intensity started to decrease, to reach at approximately $t=9$ min. a minimum of about  74\% of the initial value.  Further incubation over 70 minutes affected only weakly the SAXS profile. We observed a very slight shift of the forward scattering intensity, reaching rapidly a plateau at $\sim 15\%$ above the initial value, indicating that the micelle-monomer equilibrium is only weakly modified by the creation of a small number of aggregates. After incubation, the profile of sample cooled back to at 25 $^{\circ}$C was almost the same as the profile of the fresh sample. Slight difference between forward intensities of the initial and the incubated samples is likely due to a small number of irreversible aggregates.

%*********************************************************************  
%*********************************************************************

%*********************************************************************  
%*********************************************************************

%*********************************************************************  
%*********************************************************************

\section{Model-free interpretations}
\label{Model-free interpretations}

We first analyse the SAXS data from fig.\ref{f4} in terms of linear combination of the first (4 min.) and last (1440 min.) profiles. 
On the other hand, the last profile corresponded to $\approx 65\%$ of micelles and $\approx 35\%$ of fibrils, (neglecting the monomers) as measured by centrifugation and confirmed by analysis of SAXS profiles in term of calculated spectrum for fibrils (cf. Section \ref{Modelling}).

\begin{figure}[t]
  \centering
  \includegraphics[width=9cm]{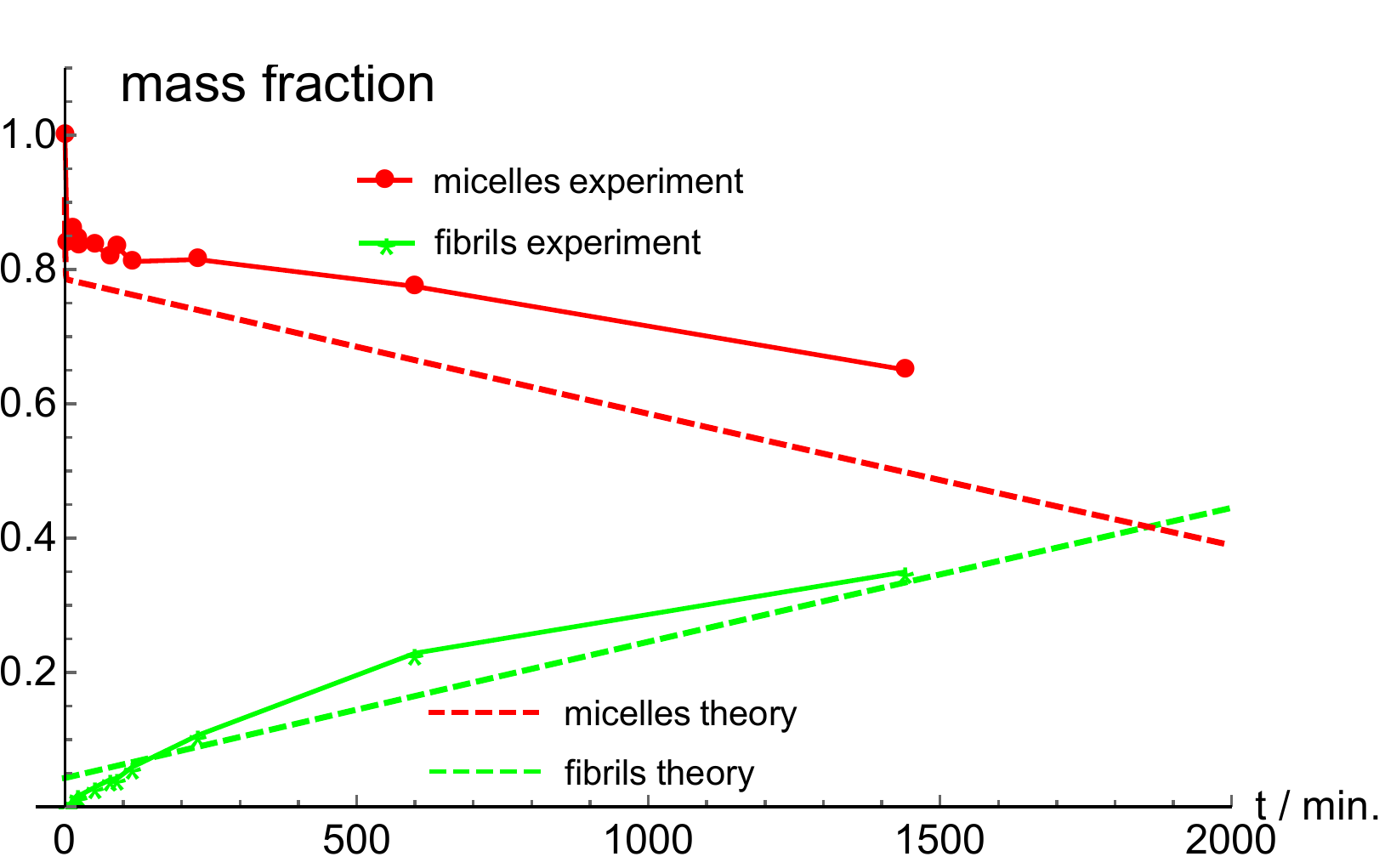}
  \caption{Fit of SAXS profiles from fig.\ref{f4} in terms of micelle and fibril mass fractions. Kinetics profiles calculated using our model (eqs. \ref{rateeqs}) are depicted in dashed lines.}
  \label{f6}
\end{figure}

The 2-components fitting procedure yielded the result shown in fig.\ref{f6}. 
The fibril mass increases approximately in a linear manner, in accordance with data in literature for 
A$\beta$ system, in which oligomer-monomer equilibrium keeps the monomer concentration constant during the fibril growth \cite{Lomakin_1997}. 

%*********************************************************************  
%*********************************************************************

The forward intensity variation from fig.\ref{f5} can be interpreted in terms of monomer release and uptake events upon heating and cooling respectively. 
In the beginning, all proteins are either in micelles or monomers, while at the end only a fraction $\eta$ of the protein remains out of fibrils. We calculated the mass loss from aggregates upon initial protein release and compared it to the mass uptake into micelles after cooling back to 25$^{\circ}$C. 
The mass release form reversible aggregates upon T-stress is approximately given by:
\begin{equation}
\Delta = \text{CMC}_{\text{h}}-\text{CMC}_{\text{c}} \; ,
\end{equation}
where $ \text{CMC}_{\text{h}}$ and $\text{CMC}_{\text{c}}$ are CMC's of the "hot" and "cold" sample respectively. This decomposition is valid and independent of 
$\eta$ as long as micelles are dominant over monomers, i.e. as long as $ \eta [prot] > CMC_c$. 
If we are dealing with micelle decay, the total micelle mass loss is given solely by decrease of the average micelle's molar mass ${\cal M}={\cal M}_0 \bar{m}$. On the other hand, the forward scattering intensity, eq. (\ref{scatintens}), is proportional to $c \bar{m}$, where $c$ is the mass concentration of scatterers, here micelles. When releasing monomers, the mass concentration of scatterers is reduced by the same factor as $\bar{m}$.
Altogether, immediately after the initial decrease of intensity, $I(0)/I(4')=(M(0)/M(4'))^2$, where $I$ and $M$ stands for forward intensities and total mass in fibrils, with arguments corresponding to time in minutes. Consequently, the fraction of released mass from micelles upon initial T-jump is $M_{\text{released}}/M(0)\approx \sqrt{I(0)/I(4')}-1\approx \frac{1}{2}\Delta I_r/I(0)=\frac{1}{2}0.095/0.60=8 \%$.
%The initial decrease of the forward intensity was 0.095$\pm 0.01$ cm$^{-1}$, while the final increase upon re-cooling was 0.21$\pm 0.02$ cm$^{-1}$. 
The intensity increase after cooling down the sample incubated for 1440 min. is $\Delta I_u= 0.21\pm 0.02$ cm$^{-1}$=0.35 $I_0$. This gives 
$M_{uptake}/M_{0}\approx \frac{1}{2}\Delta I_u/I(0)=17.5 \%$, about twice the initial release. 
In consistency with the reversibility assessment by DLS fig. \ref{f3}, the uptake of protein back to micelles is shown by SAXS to be higher as the fibrillation goes on, i.e. the scattering difference between the native state and the T-stressed state is larger if some fibrils are already formed. 
We can conclude that the effective CMC of RCMK upon T-stress is shifted to some rather high value, larger than the CMC at 25$^o$C, about 0.5g/L.
In the case of the EGCG-protected micelles, the micelles dissolution effect impacts the forward scattering intensity by $\delta I /I(0)=0.26 \pm 0.02$, confirming the LS result. Assuming that protein release is only due to micelles dissolution, we get  
$M_{released}/M(0)\approx \frac{1}{2}\Delta I_r/I(0)=\frac{1}{2}\Delta I_u/I(0)=13 \%$.

We can understand the above result if we assume that in stressed conditions the free protein, at concentration $\text{CMC}_{\text{h}}$, is composed of two conformationally different states. First is the native state that keeps the equilibrium with the micelles. The second is the conformationally modified state of protein, unable to re-integrate the micelles, but existing in equilibrium with the free native population. Thus, in stressed condition, the total concentration of monomeric protein is
\begin{equation}
 \text{CMC}_{\text{h}}\approx y_0+y_1 \; ,
\end{equation}
where $y_0$ and $y_1$ are the concentrations of monomeric protein in native and in modified state respectively. Their equilibrium is determined by
a constant $y_1/y_0=k_{\text{on}}/k_{\text{off}}$, which can be large, and depends on the degree of stress.  Consequently, in stressed condition we can have higher concentration of free protein than at 25$^o$C, contrary to the case of $\beta$-casein. Thus, the conversion of monomers to pro-amyloid conformation, not tolerated in micelles, leads naturally to the apparent increase of the effective $CMC_{\text{h}}$. The effect is present in both cases with and without EGCG, indicating that the mechanism by which EGCG inhibits fibrillation is to block the association of amyloid-prone protein into fibrils without affecting the formation of these amyloid-prone monomers and without preventing monomer release from the micelles.

The point which remains to be clarified is the increase of release/uptake effect with increasing amount of formed fibrils. 
The experiment with EGCG confirmed that the fibrils are indeed the ones that promote the effect. 
In our interpretations, the concentration of stress-induced monomeric species increases as the fibrillation goes on. 
Keeping in mind that the monomeric species are dominated by pro-amyloid monomers, this means that the presence of fibrils boosts the native-to-misfolded conversion, that is, the rate constant $k_{\text{on}}$ is an increasing function of fibrils concentration: the presence of fibrils catalyses the misfolding, which is a well known phenomenon in the context of amyloid aggregation\cite{Cohen_2011}.

The characteristic time scale of the release is of the order of 2-4 minutes. In our picture this time corresponds to single-protein release from the micelle. The single-protein exchange time of the same order was reported in 
\cite{Ma_2012} for the case of tannin-driven $\beta$-casein micelles reorganisation, which conforts our interpretation. 

Finally, altogether our data indicate that the tannin is not tolerated in fibrils. Namely, irrespective of the concentration of tannins in the explored experimental range, we found no indication of tannin-bridged fibril aggregates which suggest that in the fibrils tannin-binding sites are protected.
This result is very important since it allowed us to decipher quantitatively how tannins inhibit fibrillation, c.f. infra.

\section{Modelling}
\label{Modelling}

\subsection{Fitting SAXS form factors}

{\em RCMK alone.}
The SAXS profile before incubation $I_{0'}(Q)$ was fitted to the Svaneborg-Pedersen formula (\ref{SP}) including polydispersity (\ref{polydisp}). Best fit for the core radius was $R_0=6.25$ nm, for the radius of gyration of the Gaussian segments in the crown was $R_g=$2.8 nm, the rigid rod crown segments $l=$0.6 nm, the hydrophilic to hydrophobic ratio $\alpha=$0.7, the aggregation number $m=23$ and the polydispersity for the $R_0$ was Gaussian with $\sigma$=1.5 nm. These results are in fair agreement with data in the literature \cite{Farrell_2003,Leonil_2008} and with our DLS experiments. 

The SAXS profile at $t=4$ min. was fitted to the same model as the fresh sample $I_{0'}(Q)$, keeping all parameters fixed except the aggregation number, which decreased to $m=20$. The relative decrease of the aggregation numbers of fresh and 4 minutes samples $\Delta m/m\approx 13\%$, of the same order as the result found by naive analysis of the forward intensity. This indicates that the initial decrease of the forward intensity is due to the decrease of the aggregation number and not due to the decrease of the number of micelles.

\begin{figure}[t]
  \centering
  \includegraphics[width=9cm]{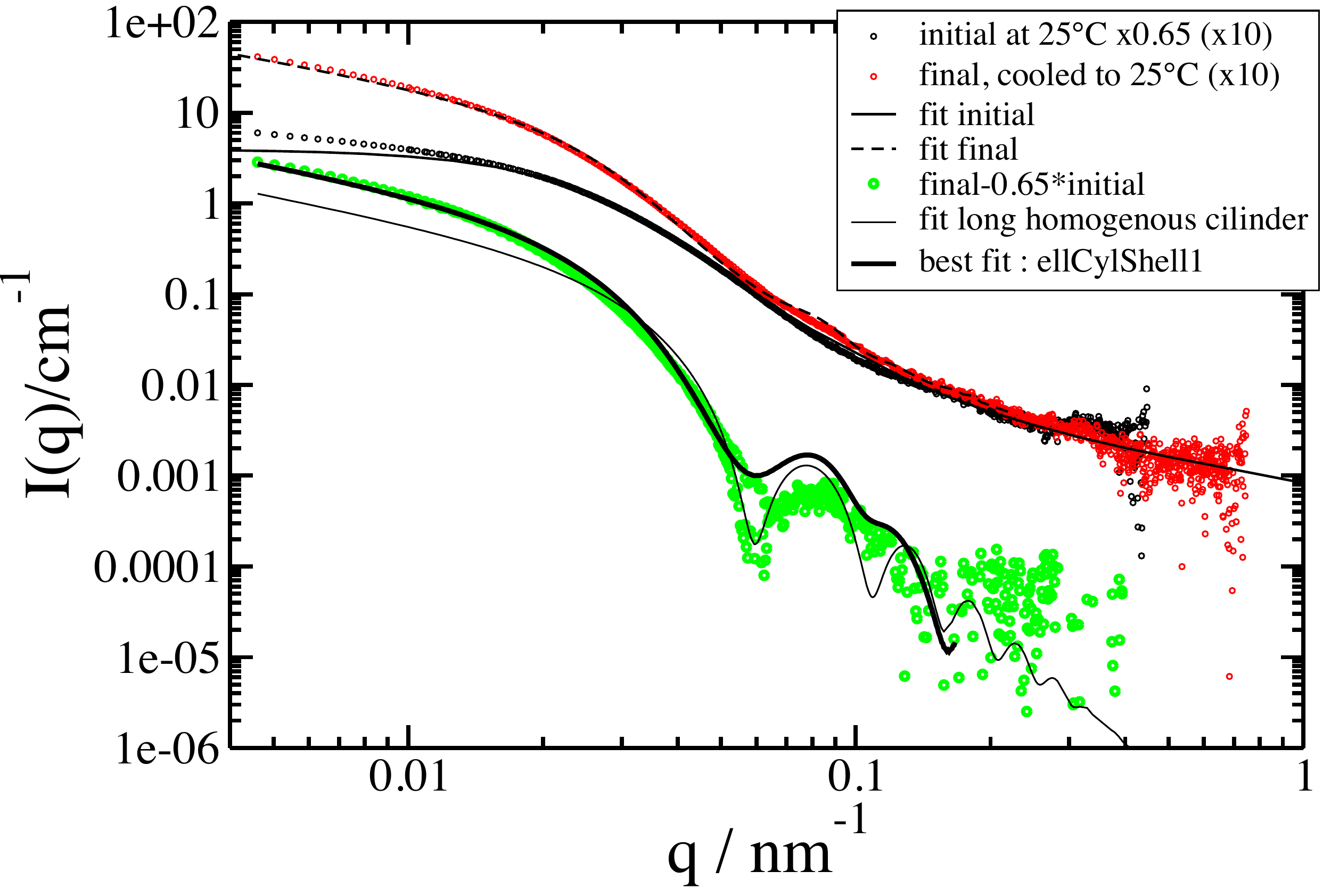}
  \caption{Identification of incubated SAXS scattering profile as superposition of micelles and fibrils, with corresponding fits.}
  \label{f7}
\end{figure}

The profile of fibrils can be found by subtracting from $I_{\mbox{final}}$ the contribution of micelles. In the present case the fraction of fibrils in the final sample was approximately 35\%, as measured by centrifugation, so we assumed that the fraction of micelles was at most 65\%.
The spectrum for fibrils $I_{fib}(Q)$ obtained as $I_{\mbox{final}}(Q)-0.65I_{\mbox{initial}}(Q)$ is shown in  fig.\ref{f7}. The data were fitted by three different models for long cylinders. The best fit was the elliptic shell cylinder model from SASFIT package \cite{SASFIT}: inner cylinder shorter radius was 2.9 nm, the shell thickness 6.0 nm, long to short axes ratio was 1.56 and the excess electron density of the core was 3.7 times higher than the excess electron density of the shell. The lengths of the fibrils could not be captured by the Q-range used in this experiment.
Above results are in good agreement with data in literature \cite{Farrell_2003,Leonil_2008}.  
%*********************************************************************  
%*********************************************************************

{\em RCMK with EGCG.}
The profile of the fresh RCMK + EGCG sample was fitted to the modified 
Svaneborg-Pedersen formula, (\ref{SP}-\ref{polydisp}), allowing for some additional contrast $\phi$ in the core.  We obtained the following fitting parameters:
the core radius $R_0=6.7$ nm, the radius of gyration for the Gaussian segments of the crown $R_g=2.5$ nm, the length of the rigid rod segments $l=0.5$ nm, the hydrophilic to hydrophobic ratio  $\alpha=0.7$, the aggregation number $m=42$, the polydispersity for the $R_0$ was Gaussian with $\sigma=0.8$ nm and the parameter $\phi=0.33$. Although the core radius of $R_0=6.7$ nm is slightly larger than the core radius 6.25 nm for the case of only RCMK, the micelle core density (\ref{exdens}) is by a factor of approximately 2.1 higher in the present case.  
The  increase of $I(Q=0)$ with respect to RCMK only is due to both the increase of the aggregation number $m$ and of the parameter $\phi$. 
Interestingly, the sample with EGCG is less polydisperse than the sample with only protein ($\sigma=0.8$ nm instead of 1.5 nm). This low polydispersity is revealed by a kink at $Q=0.09$ nm$^{-1}$.
The spectrum at 9 minutes was fitted by the same model as the previous one, modifying essentially $m$ and $\phi$. Their new values were $m=37$ and $\phi=0.21$. ($R_0$ was also reduced to 6.4 nm and the polydispersity 
increased to $\sigma$=0.9 nm.)  These results reveal that the protein-tannin complex micelles behave in a similar way as simple protein micelles upon heating: the individual proteins (probably coated with tannin) are released from micelles, without creating alternative non-fibrilar structures. 
In particular, we see no indications of fragmentation into smaller micelles with $m>1$. 
The fraction of released tannin-protein complex from the micelles $\Delta m/m =M_{released}/M_{0} \approx (42 - 37)/37=14 \%$ matches the result obtained from forward intensity analysis without structural assumptions.

\subsection{Theoretical modelling of kinetics and implementation of the effects of EGCG}

In what follows we construct a simple kinetics model based on the assumption that the fibrils grow by amyloid-prone monomer accretion  
onto oligomeric seeds. We suppose that these oligomeric seeds are generated in micelles. 
Furthermore, we suppose that the amyloid-prone monomers involved in the accretion are formed by rapid conversion of free monomers. 
Our goal was to find out if this model can reproduce the initial decrease of LS and SAXS scattering intensity  (without affecting the effective hydrodynamic diameter $D_H$), and the subsequent growth profiles, as shown on figs. \ref{f2}, \ref{f6} and \ref{f7}.

Inspired by the Lomakin's approach \cite{Lomakin_1997}, we construct the rate equations for concentrations of free native monomers $y_0$, free pro-amyloid monomers $y_1$, monomers in micelles $M$, monomers in fibrils $N_1$ and number of fibrils $N_0$ at given total protein concentration [prot] (see "Methods" section for definitions):
\begin{subequations}
\begin{align}
 \frac{dy_0(t)}{dt}&= k_{\text{off}}y_1(t)-k_{\text{on}}y_0(t)+k_{\text{mic}}\times \nonumber \\
 &\times [M(t) -  M_0([prot] - N_1(t) - y_1(t))] \label{rateeqsa}
 \\
  \frac{dy_1(t)}{dt} &= -k_{\text{off}}y_1(t)+k_{\text{on}}y_0(t)-k_{\text{e0}} f_{\text{inh}}(t) y_1(t)N_0(t)
  \label{rateeqsb}
  \\
\frac{dN_0(t)}{dt} &= 0 \label{rateeqsc}
\\
  \frac{dN_1(t)}{dt} &= k_{\text{e0}} f_{\text{inh}}(t) y_1(t)N_0(t)     \label{rateeqsd}
  \\
  M(t) & = \text{[prot]} - y_0(t) - y_1(t) - N_1(t) \label{rateeqse} \; ,
  \end{align}
  \label{rateeqs}
  \end{subequations}
where the last equation is just the total protein conservation. 
Reversible conversion of monomers from native ($y_0$) to pro-amyloid conformation ($y_1$) is given by the first two terms of eqs. (\ref{rateeqsa}) and (\ref{rateeqsb}). The third term in eq. (\ref{rateeqsa}) is the release of material from out-of-equlibrium micelles $M$. The release rate is proportional to the 
shift between actual $M$ and the equilibrium $M_0([prot] - N_1(t) - y_1(t)))$, calculated at actual concentration of protein in native conformation only.  We assume that the  equilibrium function 
$M_0(x)$ is given by the shell model \cite{Mikheeva_2003}. The third term in (\ref{rateeqsb}) and the eq. (\ref{rateeqsd}) is the accretion of fibrils by addition of single pro-amyloid monomers onto the ends of existing fibrils.
The growth is controlled by the rate constant $k_{e0}$, the only process modified in the presence of tannins. We model this modification by the inhibition function $f_{\text{inh}}(t)$.
In constructing the function $f_{\text{inh}}(t)$, we assumed: (i) it is monotonically decreasing function of the tannin-to-protein molar ratio $\theta \equiv $[tan]/[prot] and (ii) the tannins are not tolerated in fibrils, so that the effective  tannin-to-protein molar ratio $\theta_{\text{eff}}$ increases as the fibrillation goes on:
we put  $\theta_{\text{eff}}$=\text{[tan]}/(\text{[prot]}-$N_1)$. 
Altogether, we postulate that $f_{\text{inh}}(t)$ has a soft-threshold form:
\begin{equation}
f_{\text{inh}}(t)=f_{\text{tr.}w}\left(\theta _{\text{eff}}-\tau \right)=f_{\text{tr.}w}\left(\frac{\theta }{1-\nu}-\tau \right) \; ,
\label{finh}
\end{equation}
where $\nu \equiv N_1(t)/ \text{[prot]}$ is the molar fraction of fibrillar protein (pro-amyloid monomers are excluded) 
and $f_{\text{tr.}w}(x)$ is some soft cutoff function around $x=0$ with softness $w$.
The main reason for this choice of inhibition function is that tannin really blocks the fibrillation after some time, rather than just slowing it down. 
We assume that $f_{\text{tr.}w}(x)$ has the following form
\begin{equation}
f_{\text{tr.}w}(x)=\frac{1}{1+e^{x/w}}
\label{fd}
\end{equation}

Some of initial conditions and parameters that determine the kinetics are known, or easily determined, while some of them remain to be determined by fitting the overall kinetics profiles to kinetics data.
The parameters that are easily found by comparing final and initial scattering intensities and effective hydrodynamic radii are the number of monomers per nm of fibril length $\lambda$ and the concentration of fibrils $N_0$.
We know that $\lambda=\bar{n}(\infty)/L(\infty)$, where the aggregation number of mature fibrils $\bar{n}(\infty)$ is found approximately by assuming that in the beginning of incubation all protein is in micelles and at the end all protein is in fibrils. This implies: 
\begin{equation}
\frac{\bar{n}(\infty)f_{\mbox{\tiny rod}}\hspace{-1mm}\left(\frac{ L(\infty)}{54 \; \text{nm} }\right)}{\bar{m}(0)}=\frac{I(\infty)-I_b}{I(0)-I_b} \; . 
\end{equation}
From our LS results this ratio is approximately 10, the micelles aggregation number is $\bar{m}(0)=24$ (in our model kept constant upon incubation), and we have seen by AFM that $L(\infty)\approx 480$ nm. Altogether, we get $\bar{n}(\infty)\approx 630$ and $\lambda \approx 1.32$ nm$^{-1}$.
The concentration of fibrils (assumed to remain constant and consequently equal to the concentration of seeds) is obtained from $N_0=$[prot]/$\bar{n}(\infty)\approx 0.25\; \mu$mol/L. The size (aggregation number) of the seeds is assumed to be equal to the micellization number without tannin  $\bar{m}_0$, in agreement with the seeding mechanisms proposed by Lomakin {\em et al.} \cite{Lomakin_1997} for A$\beta$ and by Leonil {\em et al.}  \cite{Leonil_2008} for $\kappa$-casein. Namely, according to these authors the confinement of protein in micelles promotes formation of amyloid seeds, with the size comparable to micelles.

The rate of monomer release from micelles $k_{\text{mic}}(M(t) - M_0(c - N_1(t) - y_1(t)))$ was adjusted to get typical release time of the order of $2-4$ min, which, for our choice of  the function $M_0(x)$, gives $k_{\text{mic}}=5$ min$^{-1}$.
To calculate $M_0(x)$ within the the shell model  the parameters were adjusted to give a rather low CMC$_h$ of 0.16 g/L (which is fairly correct for $\beta$-casein \cite{Mikheeva_2003}) and the average micellization number $\bar{m}=24$ at 3 g/L of protein, corresponding to the one that fits SAXS data. 

The rate constants $k_{\text{on}}$ and $k_{\text{off}}$ are chosen to (i) reproduce the right magnitude of the initial monomer release, determined jointly by $k_{\text{on}}/k_{\text{off}}$ and the value of CMC$_h$, and (ii) to have rather high values because we assume that the conversions of monomers from native to amyloid-prone conformation fort and back are much more rapid than all other processes. Altogether, we put $k_{\text{on}}=5 k_{\text{off}}=1000$ min$^{-1}$.

To calculate both $I(t)$ and $D_{\text{eff}}(t)$ from the system (\ref{rateeqs}) using formulas (\ref{Isimp}) and (\ref{Rsimp}) the calibration of fibril hydrodynamic radius $R_H(L)$ was necessary. 
Since in our model the fibrils start to grow from oligomers that have the size of micelles, the calibration matching our system is well approximated by a simple form $R_{H,{\mbox{\tiny fib}}}( L)=R_{H,{\mbox{\tiny mic}}}+0.045 L$, with micelles hydrodynamic radius $R_{H,{\mbox{\tiny mic}}}$ matching the DLS data at $t=0$, and kept constant during kinetics.

The initial condition were: $y_0(0)=[prot]-M_0([prot])$, $y_1(0)=0$, $N_0=0.25\; \mu$mol/L, $N_1(0)=\bar{m}_0 N_0$. The micellization number $\bar{m}$ in presence of tannin was adjusted in a way to fit the LS scattering intensities at $t=0$. Its value varied between 24 without EGCG and 80 for 1 g/L of EGCG.

The whole series of DLS curves for EGCG concentrations up to 1 g/L, together with the fits using the above theory is shown on fig. \ref{f2}. 
The zoom to the initial protein release from micelles with and without EGCG can be seen on fig. \ref{f3}.
The best fitting value of the fibril accretions rate $k_{e0}=0.12 \times 10^6$ min$^{-1}$ L/mol for pure protein was adjusted by fitting the overall profiles for $I(t)$ and $D_{\text{eff}}(t)$.

To fit the kinetics at 40 $^\circ$C extracted from SAXS, fig.\ref{f7}, two parameters were modified: parameter $k_{e0}$ was divided by 20, while the constant $k_{\text{off}}$ was increased by a factor of 2.5. These modifications reflect the fact that the fibrillation kinetics at lower stress is slower and that conversion of monomers from native to pro-amyloid conformation is less efficient.

\begin{figure}[t]
  \centering
  \includegraphics[width=9cm]{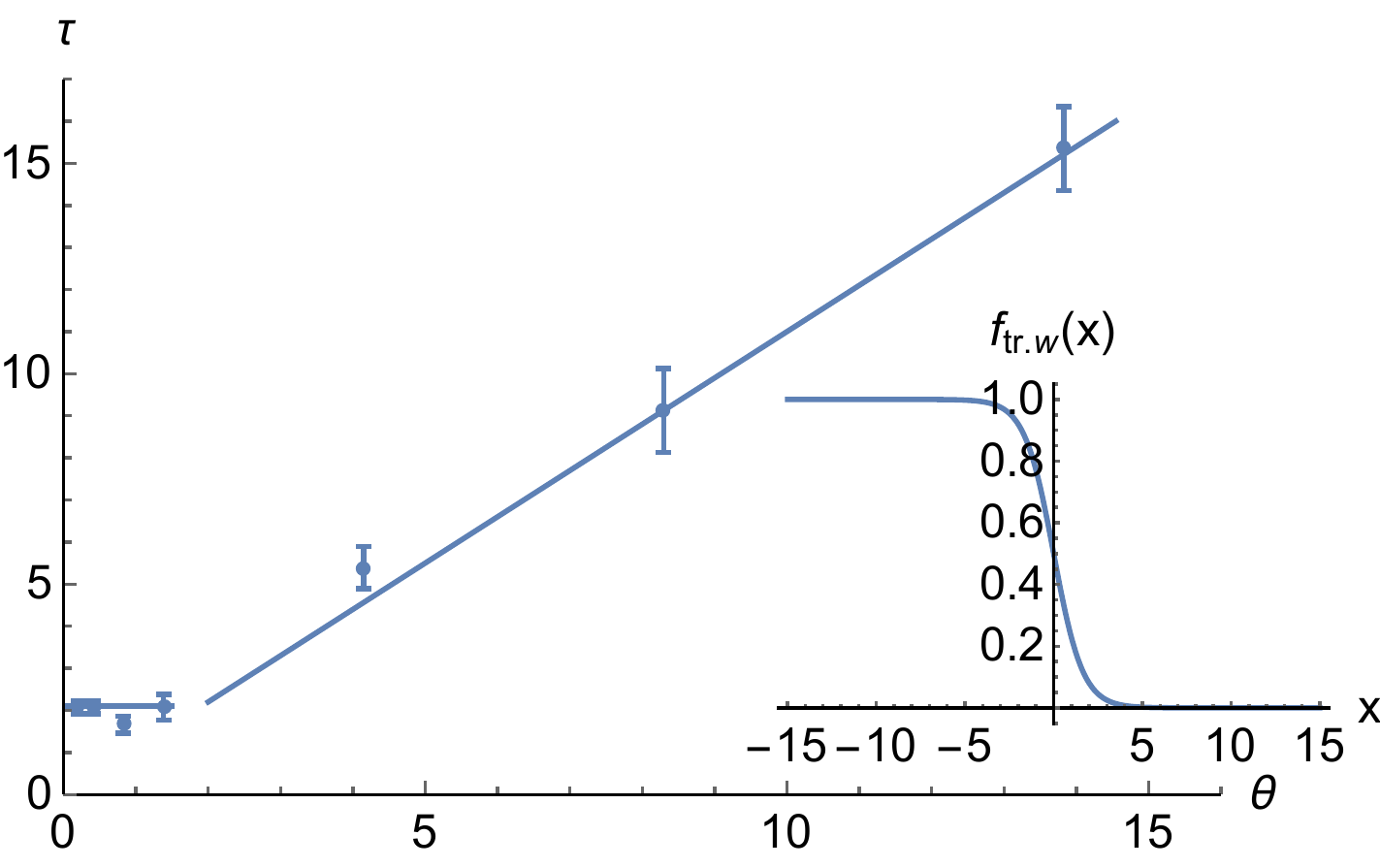}
  \caption{Inhibition parameter $\tau $ as function of tannin/protein molar ratio $\theta$ at 45$^{\circ}$C. Inset: function $f_{\text{tr.}w}(x)$ for $w=0.82$.}
  \label{f9}
\end{figure}

In fitting the kinetics data on fig.\ref{f2}, the essential fitting parameter to implement the effect of EGCG was the inhibition threshold $\tau(\theta)$ playing in the inhibition function (\ref{finh}). The function $f_{\text{tr.}w}(x)$ for $w=0.82$ is shown in inset of fig.\ref{f9}. The quantity $\tau$ is interpreted as the threshold of the effective tannin to free protein ratio $\text{[tan]}/(\text{[prot]}-N_1)$ above which the fibril accretion rate is blocked.
Choosing the softness parameter $w=0.82$, we are able to fit the whole series by adjusting $\tau(\theta)$ and the micelles scattering intensity, adjusting $\bar{m}$. The resulting fitting values of $\tau(\theta)$ are shown on fig.\ref{f9}. Two regimes can be identified. First, for $\theta \lesssim 2$ is the  threshold regime, in which $\tau=\tau_0=2\pm 0.2 $ remains constant. In the second regime, $\theta > 2$, the value of $\tau$ becomes equal to  $\theta$.

\section{Discussion}
\label{Discussion}

\subsection{A consensus with existing interpretations}
Our experimental data are reproduced by a model that assumes that the mechanism of RCMK fibrillation is the addition of single pro-amyloid monomers onto oligomeric seeds, released from micelles.
The effect of tannin EGCG is to block this accretion without affecting neither the release of monomers from micelles, nor the native to pro-amyloid conversion of monomers.
The release of monomers from the micelles even in the absence of fibrilation is interpreted by the exclusion of the pro-amyloid monomers from micellar assemblies. Non-native monomers, thereby do not participate in micelle-monomer reversible exchange.

The proposed model eqs. (\ref{rateeqs}) reconciliates two apparently opposing interpretations of the RCMK fibril formation: The Rennes group \cite{Leonil_2008} points to the conformational modification within micelles as determinant event in the RCMK fibrillation. These irreversibly modified micelles act subsequently as building blocks for fibrils, without participation of monomers. Alternative interpretation, due to Adelaide group \cite{Ecroyd_2008,Ecroyd_2010}, is that the monomer release from micelles is the rate limiting event, and that these monomers are rapidly aggregated into fibrils, keeping monomer concentration nearly zero.

In our model, both ideas are incorporated, but with some modifications. The seeds for fibril growth are indeed the modified micelles, but the building blocks are not these modified micelles but free pro-amyloid monomers. In SAXS experiment we have not detected any conformational change of the micelles, change expected if micelles were transformed into amyloid oligomers.
Moreover, if internal irreversible structure modification of micelles were the precursor of fibrillation (Rennes group hypothesis), the low-Q scattering intensity is not supposed to decrease in the first minutes of incubation, in disagreement to what happens in LS and in SAXS experiments. One could, however, imagine that the initial conformational re-arrangement invokes some decrease of micelle mass, which would explain the initial decrease of $I$. This possibility is ruled out by experiments shown in (figs. \ref{f3} and \ref{f4}), which demonstrate that the effect is reversible in the presence of EGCG, and consequently is a micelle-monomer exchange and not the irreversible transformation of micelles. In our interpretation only a small fraction (in our model $\bar{m}/\bar{n}\approx 4$ \% of micelles) are conformationally converted, implying that the effect is completely dominated by the majority made of natively conformed micelles, as we observed by SAXS. 
Alternatively, if the released monomers were immediately incorporated into fibrils, as proposed by the Adelaide group, the initial drop of $I$ could indeed be detected.
However, after cooling back to room T, $I$ would not re-increase since all material released from micelles would readily be incorporated into irreversible fibrils. 

Altogether, the two main modifications of existing models that we propose are: (i) {\em only a small umber of micelles give rise to irreversible amyloid oligomeric seeds}, with the aggregation number close to the one of micelles. These modified micelles do not aggregate together, but grow due to accretion, and (ii) monomers are indeed released from micelles prior to fibrillation, but these monomeric species are subject to conformational changes to pro-amyloid forms, and along the fibril growth constitute  {\em a pool of monomeric proteins} that keeps micelles in a quasi-equilibrium, and can get reintegrated in micelles upon re-cooling. 

In our interpretation EGCG does not inhibit conformational changes from native to pro-amyloid conformations. This might sound somewhat misleading, in particular in the light of reported CD results on RCMK  \cite{Hudson_2009} and on similar systems $\alpha$S and A$\beta$ \cite{Ehrnhoefer_2008}, where clear signatures are found that EGCG suppresses pro-amyloid conversion. The solution of this apparent contradiction is in the fact that CD detects both pro-amyloid monomers $(y_1)$ and protein within fibrils $(N_1)$, where $N_1(t)$ dominates the amyloid part of CD signal, while $y_1$ is low (in comparison to [prot]) and is roughly constant over time at constant temperature. Namely, the experiment on RCMK \cite{Hudson_2009} is performed at constant incubation temperature of 37$^{\circ}$C.
Our mechanism indeed predicts that EGCG inhibits the growth of fibrils, which reflects on decrease of CD signature of amyloid conversion. 
The effect of the increase of pro-amyloid monomeric pool $y_1$ upon heating could in principle be observed in a separate experiment in which the system is subjected to a T-jump within the CD cell, similarly to our DLS and SAXS experiment. 

Strictly speaking, our kinetics model fails to reproduce precisely all observations. The first is the increase of the monomer pool ($\approx $CMC$_h$) upon fibril growth. It is not reproduced since we assumed that conversion of monomers from native to pro-amyloid conformation is constant. To remedy this point, the ratio $k_{\text{on}}/k_{\text{off}}$ should be an increasing function of $N_1$, which is a way to implement the self-catalysis of amyloid species by the presence of already formed fibrils. That way the concentration of pro-amyloid monomers $y_1$ would increase along the fibril growth, keeping $y_0$ approximately constant. The  part of the monomer pool taking part in the temperature triggered release-uptake (CMC$_h$-CMC$_c\approx y_1$) would thereby increase, as observed (figs. \ref{f3} and \ref{f5}). 

Second limit of the present model is that the calculated growth profiles do not fit exactly the experimental data for from DLS (fig.\ref{f2}) and SAXS (fig.\ref{f7}).
One reason is because
polydispersity effects on the observables $I(t)$ and  $D_{\text{\tiny eff}}$ are not taken into account since the rate equations are established only for first two moments of micelle and fibril distributions $m_i$ and $n_i$, see (\ref{LSsub}) and eqs. (\ref{rateeqs}). Secondly, the accretion rate $k_{e0}$ is kept constant independently of the fibril size and age. Experimental data are somewhat steeper than the calculated profiles in the beginning of the kinetics, and this shift gets less pronounced as aggregation progresses. To cure this, one could think of incorporating some $N_1$ dependence in $k_{e0}$, in a way that older are the fibrils, less efficient is the monomer addition. The effect could be explained by taking into account the decrease of the fibril-monomer encounter frequency due to decrease of diffusion with increasing fibril length.

\subsection{Inhibition function and stress magnitude}

The effects of tannin on fibril growth was incorporated in our model in a semi  phenomenological  way by introducing the inhibition function $f_{\text{inh}}(t)$ as a multiplier for $k_{e0}$. The $t$-dependence in  $f_{\text{inh}}(t)$ is completely determined by the concentration of fibrillar protein $N_1(t)$, see eq. (\ref{finh}).
Very important experimental fact used in construction of  $f_{\text{inh}}$ is that tannin EGCG indeed {\em blocks} the fibril growth at some finite, [tan]-dependent level of fibrillation, in opposition to some other molecular chaperones that delay the fibrillation rather than block it \cite{Arosio2016}.
The basic property of EGCG allowing for this blocking effect is that EGCG does not participate in formed fibrils, but inhibits the accretion by preventing addition of individual pro-amyloid protein onto fibril ends. 
Therefore, at some intermediate tannin-to-protein ratio, along with the formation of fibrils, the effective tannin concentration in regard to non-amyloid protein increases, which furthermore inhibits the growth rate of remaining protein and finally blocks fibril accretion completely above some threshold tannin-to-free protein molar ratio $\tau _0$. The fit of the inhibition threshold $\tau$ as function of total tannin-to-protein molar ratio $\theta$ indicates that for $\theta \lesssim \tau_0$ we expect some finite growth before final fibrillation blocking. For our conditions of strong stress (T=45$^\circ$ C) the characteristic value of $\tau _0$ is rather high, of the order of 2. At lower stress (T=37$^\circ$ C) the fibrils start to be substantially blocked already at $\tau _0\approx 0.3\pm0.1$ \cite{Hudson_2009}. 
In fact, the tannin dependence of the kinetics is completely determined by the value of $\tau _0$, itself a rather steep function of the temperature shift from 25 $^\circ$ C.
To make it obvious, we notice first that the dependence $\tau (\theta)$ can be approximated by $\tau \approx \sqrt{\tau _0^2+\theta ^2}$. Inserting it in the expression (\ref{finh}) and using eq. (\ref{fd}) we get:
\begin{equation}
f_{\text{inh}}=\frac{1}{1+e^{\left[\frac{\tilde{\theta}}{1-\nu(t)}-\sqrt{1+\tilde{\theta}^2}\right] /\tilde{w}}} \; .
\end{equation}
To reproduce the experiment, the same values of reduced variable $\tilde{\theta}\equiv \theta/\tau_0$ must give the same degree of blocking, and, after rescaling the time and appropriate adjustment of $k_{\text{on}}/k_{\text{off}}$, the same growth profiles. This implies that $\tilde{w}\equiv w/\tau_0$ must be constant. It is found from the fitting values at 45$^\circ$ C that $w=0.82$ and $\tau _0=2$ giving $\tilde{w}=0.41$ which is the universal softness parameter.
Altogether, we see that $\tau _0(T)$ is a way to measure the magnitude of amyloid stress at given temperature. Once  $\tau _0$ is known, kinetics at any EGCG concentration can be predicted using our simple theory.

In constructing inhibition function, no particular assumption was made about EGCG-protein affinities or about microscopic mechanisms by which EGCG prevents fibril accretion. However, we have seen that EGCG is not tolerated in fibrils and that it is typically bound to hydrophobic portions of the chain, as it can be concluded from SAXS results. Namely, in normal conditions the majority of bound tannin are in the core of micelles, as revealed by the increase of the parameter $\phi$, see 
\ref{Modelling}. Most of hydrophobic protein segments being gathered in micelles core, we conclude that these are also preferential sites for tannin binding.
This effect has been reported for $\beta$-casein \cite{Zanchi_2008epl,Ma_2012} or even for salivary proteins \cite{Zanchi_2008jp}. 
These facts, together with the arguments on non-specific and mostly hydrophobic nature of EGCG-protein binding in the context of amyloid protection based on NMR \cite{Hudson_2009,Ehrnhoefer_2008},
provide some hints for constructing molecular-based mechanisms. In particular, one can speculate that the degree of protection is given by the adsorption of EGCG on exposed $\beta$ strands at the ends of fibrils. For low stress (or low $\tau_0$) only very few EGCGs should suffice to prevent amyloid accretion, while at higher stress more EGCGs are needed. However, details and quantitative microscopic description of the effect remains out of the scope of the present analysis.

\subsection{Tannin-induced compaction and reversible tannin-protein colloidal complex}
Both DLS and SAXS data demonstrate that at room temperature micelles become heavier and more compact when EGCG is added. The analysis of SAXS form factors indicate that the increase of micelle's density is due to both increment of average micellization number $\bar{m}$ and the change in contrast of the micelle core $\phi$, due to both EGCG and protein intake. Similar effect was reported for the case of $\beta$-casein with EGCG \cite{Ma_2012}.
These effects are incorporated in our present model in a somewhat artificial way by high values of micellization number, up to $\bar{m}=80$,
needed to fit the DLS data.
It exceeds the value $\bar{m}=42$ found by structural fits of SAXS data, but matches well the modification of SAXS forward intensity: it increased by a factor of 2/0.6= 3.3, which is equal to the increase of the LS scattering intensity $80/24$. 
This discrepancy between the structural data and values obtained by simple comparison of the scattering intensities can be understood if we keep in mind that EGCG contributes the scattering intensity under the form of complexes with the protein and that it brings additional protein in the core of micelles, which is rationalized over the parameter  $\phi$, see (\ref{Structural}) and (\ref{Modelling}).

The most intriguing result in this context is that T-stress induces reversible release of material from micelles. It is plausible to assume that these released species are protein monomers complexed to EGCG, similarly to the case of $\beta$-casein \cite{Ma_2012}. However, the reversibility of the micellization of this complex was not obvious.
To our knowledge, the T-triggered micelle-monomer conversion of EGCG-protected RCMK system is the first example of completely reversible tannin-protein complex.

%*********************************************************************  

%*********************************************************************

\section{Conclusions}
\label{Conclusions}

The first objective of this work was to decipher the mechanisms of RCMK fibril formation, on the basis of newly found effect: the release of free monomers from the micelles upon thermal stress. Second objective was to incorporate in these mechanisms the effects of  tannin EGCG, nowadays one of the most efficient inhibitors of amyloid fibrillation.
Important conclusions are accessible already at the "model-free" level, Section \ref{Model-free interpretations}. With no need for detailed structural assumptions on micelles 
and without assuming any specific form of rate equations for fibrillation, DLS and SAXS experiments enable to conclude that i) the thermal stress increases the effective CMC, ii) the micelles release a fraction of their mass, and iii) this fraction increases as the fibrillation goes on.  
Experiments in the presence of EGCG revealed even stronger stress-triggered monomer release from the micelles. The monomer release was in all cases reversible. Furthermore, we found no indication of tannin bound to fibrils.  

On the basis of these facts and of structural information from fits of SAXS spectra
we propose a simple kinetics model, reproducing quantitatively the overall kinetics profiles. In particular, within our model  
(i)the initial low-Q scattering intensity drop due to monomer release is reproduced
(ii) the blocking effect of EGCG is understood, because kinetics at long times reach the plateau which decreases as EGCG concentration increases.
All effects of EGCG are incorporated in the inhibition function for the fibril accretion  rate. In our model the EGCG concentration is compared to a characteristic parameter $\tau _0$, the critical tannin-protein concentration ratio needed to block fibrillation. Thus, for a given level of thermal stress, parametrized by  $\tau _0$, kinetic profiles at any tannin concentration can be predicted.

% Propensity of $\kappa$-casein to form fibrils is believed to be related to presumably horse-and-rider-like conformation \cite{Kumosinski_1993} of native monomer in which the beta-strands are fairly exposed. 

 %Amyloid Fibril Networks in context of functional materials with physiological functions in living organisms, but also in artificial variants. Elasticity in Physically Cross-Linked Amyloid Fibril Networks \cite{Cao:2018aa}
 
   %fast AFM  In Situ Observation of Amyloid Nucleation and Fibrillation by FastScan Atomic Force Microscopy \cite{Huang_2018}

\section{Acknowledgments}
We are indebted to G\'eraldine Hallais, Herv\'e Aubin and Michael Rosticher for help with AFM experiment, Pierre Roblin and Javier P\'erez for technical support with SAXS experiment, Jan Skov Pedersen and Tommy Nylander for advices concerning micelle form factors, and John A. Carver for a precious discussion concerning carboxymethylation of $\kappa$-casein. The synchrotron facility SOLEIL is acknowledged for financial support and the provision of beam time. DZ and CT acknowledge support by Programme Investissement d'Avenir ANR-11-LBX-0011.

%\section{Supporting Information Available} 


%\begin{figure*}[p]
%\centering
%  \includegraphics[height=3.4375cm]{TOC.pdf}
 % \caption{TOC graphics}
  %\label{TOC}
%\end{figure*}

\end{document}